\documentclass[12]{article}
\usepackage{amsmath,color,xcolor}
\usepackage{graphics,graphicx,amsbsy,amssymb}

\newcommand{\JJ}{{\boldmath \mbox{$J$}}}

\newcommand{\uu}{{\boldmath \mbox{$u$}}}

\newcommand{\XX}{{\boldmath \mbox{$X$}}}

\newcommand{\rr}{{\boldmath \mbox{$r$}}}

\newcommand{\vv}{{\boldmath \mbox{$v$}}}

\newlength{\defbaselineskip}
\newcommand{\setlinespacing}[1]%
           {\setlength{\baselineskip}{#1 \defbaselineskip}}

\begin{document}

\title{\textbf{ Externally driven macroscopic systems: Dynamics versus Thermodynamics}}
\author{Miroslav Grmela\footnote{ e-mail:
miroslav.grmela@polymtl.ca}\vspace{0.5cm}\\
\'{E}cole Polytechnique de Montr\'{e}al,\\
  C.P.6079 suc. Centre-ville
 Montr\'{e}al, H3C 3A7,  Qu\'{e}bec, Canada}
\maketitle

\begin{abstract}

Experience collected in mesoscopic dynamic modeling of externally driven systems indicates absence of potentials
that could play role of equilibrium or nonequilibrium  thermodynamic potentials yet their \\ thermodynamics-like  modeling is often found to
 provide a good description, good understanding,  and predictions that agree with results of experimental observations. This
 apparent contradiction is explained by noting that the  dynamic and the thermodynamics-like investigations on a given mesoscopic level
 of description are not directly related. Their relation is indirect. They both represent
two aspects of  dynamic modeling  on  a more microscopic level of description. The thermodynamic analysis arises in the investigation
of the way the more microscopic dynamics reduces to the mesoscopic dynamics (reducing dynamics) and the mesoscopic dynamic analysis in  the
investigation of the result of the reduction (reduced  dynamics).

\end{abstract}

\section{Introduction}\label{Intr}

Boussinesq equation is  a well known  example of   mathematical formulation of mesoscopic dynamics of externally driven macroscopic systems. The mesoscopic level on which the physics is regarded in this example is the level of fluid mechanics,
the system itself is  a horizontal layer of fluid heated from below (Rayleigh-B\'{e}nard system), and the external driving forces are the gravitational force and  imposed temperature gradient.  Analysis of solutions of the Boussinesq  equations reveals properties observed in experiments (e.g. the observed
passage from less organized to a more organized behavior presents itself as a bifurcation in solutions). Many other examples of this type can be found for instance in \cite{Halp}. One of the common features of the dynamical equations that arise in the examples (the feature that has been noted in \cite{Halp}) is that  there does not seem to be possible, at least in general, to associate them
with  a potential  whose landscape would provide a pertinent  information about their solutions [\textit{ "... there is no evidence for any global minimization principles controlling the structure ..." - see the last paragraph of Conclusion in \cite{Halp}}].  Since potential (or potentials) of this type are essential in any type of thermodynamics, the observed common feature seems to point to the conclusion that  there is no thermodynamics  of externally driven systems.

On the other hand, there is a long tradition (starting  with Prigogine in \cite{Prigogine}) of  investigating  externally driven systems with methods of thermodynamics.
Roughly speaking, responses of macroscopic systems to external forces are seen as adaptations  minimizing their resistance.  The thermodynamic potentials involved in this type of considerations (i.e.  potentials used to characterize the "resistance") are usually various versions  of the work done by external forces and the  entropy production. There are many examples of very successful and very useful considerations of this type  (see e.g. \cite{Umb}). In Section \ref{EX5} we illustrate the  thermodynamic analysis in the context of an investigation of the  morphology of immiscible blends. Specifically, we show how the thermodynamic argument  provides an estimate of concentrations at the point of phase inversion, i.e. at the point at which the morphology of a mixture of two  immiscible fluids  changes in such a way that the roles of being encircled and encircling  changes (i.e. the continuous phase and the dispersed phase exchange their roles).

The experience  collected in investigations of externally driven systems can be thus summed up  by saying that mesoscopic
dynamical modeling indicates an impossibility of  using thermodynamics-like arguments yet  this type of arguments are often found to be very useful and pertinent. There are in fact well known examples \cite{Keizer}  in which both dynamic and thermodynamic approaches were developed and the potentials used in the thermodynamic analysis are proven  to play no significant role in  the dynamic analysis.
Our objective in this paper is to suggest an explanation of this apparent contradiction. We show that the dynamic and the thermodynamic analysis made on a given mesoscopic level of description are not directly related. Their relation is indirect.  They are both two aspects of a single dynamic analysis  made on a more microscopic (i.e. involving more details) level of description. An investigation of the way the microscopic dynamics is reducing to the mesoscopic dynamics  provides the  mesoscopic thermodynamics (Section \ref{RD}) and the investigation of the final result of the reduction provides the mesoscopic dynamics.

It is important to emphasize that we are using in this paper  the term "thermodynamics"  in a general sense (explained in Section \ref{RD}). While the  classical equilibrium thermodynamics and the Gibbs equilibrium statistical mechanics are particular examples of the general thermodynamics presented in Section \ref{RD}, they are not the ones  that are the  most pertinent for discussing externally driven systems.

\section{Multiscale Mesoscopic Models }\label{MMM}

Given an externally driven system (or a family of such systems), how do we formulate its dynamical model?
The most common way to do it (called hereafter a direct derivation) proceeds in the following three steps. First, behavior of the externally driven macroscopic systems under consideration  is observed experimentally in certain types of measurements called hereafter \textit{meso-measurements}. In the second step, the experience collected in the  meso-measurements together with  an insight into the physics taking place in the observed systems leads to the choice of the level of description, i.e. the choice of state variables (we shall denote them by the symbol $y$), and equations
\begin{equation}\label{Gdyn}
\dot{y}=g(y,\zeta, \mathcal{F}^{meso})
\end{equation}
governing their time evolution. By  $\zeta$ we denotes the material parameters (i.e. the parameters through which the individual nature of the physical systems under consideration is expressed) and $\mathcal{F}^{meso}$ denotes the external forces.  In the third step,  the governing equations (\ref{Gdyn}) are solved and the solutions are compared with results of observations. If the comparison is satisfactory, the model represented by (\ref{Gdyn}) is called a well established mesoscopic dynamical model (e.g. the Boussinesq model is a well established model of the Rayleigh-B\'{e}nard systems).
The choice of state variables $y$ in the second step is usually made   by trying to formulate the simplest possible model in the sense that the chosen state variables are related as close as possible to the quantities observed in the \textit{meso} measurements. The original derivation of the Boussinesq equations constituting the dynamic model of the Rayleigh-B\'{e}nard system provides a classical example of the direct derivation. The chosen mesoscopic level is in this example the level of fluid mechanics (the classical hydrodynamic fields serve as state variables $y$). The comparison made in the third step shows indeed agreement between predictions of the model and results of experimental observations.  Hereafter, we shall refer to the collection of \textit{meso} measurements and the mathematical model (\ref{Gdyn}) as a \textit{meso level} description.

 We now pick one well established mesoscopic model (e.g. the Boussinesq model).
There are  immediately two conclusions that we  can  draw. The first one is that there exist more microscopic levels (i.e. levels involving more details, we shall call them \textit{MESO levels})  on which the physical system under investigation can be described. This is because the  chosen  \textit{meso level} (e.g. the level of fluid mechanics) ignores many microscopic details that appear to be irrelevant to our interests (determined by meso-measurements  and also by intended \textit{meso} applications).  We recall that there always exists at least one well established \textit{MESO level} on which states  are described by position vectors and velocities of
$\sim 10^{23}$ particles composing the macoscopic systems under consideration (provided we remain in the realm of classical physics). Such ultimately microscopic model will be hereafter denoted as \textit{MICRO} model.

The second conclusion is that
if we   choose a \textit{MESO level} and we found it to be well established (i.e. its predictions agree with results of more detailed \textit{MESO} measurements),  then we have to be able to see in solutions to its governing equations the following two types of dynamics:  (i)   reducing dynamics describing  approach to the \textit{MESO} dynamics to the \textit{meso} dynamics, and (ii)  reduced \textit{MESO} dynamics that is the \textit{meso} dynamics.  This is because both the original \textit{meso}  model and the more microscopic \textit{MESO} model have been found to be well established. Following further the second conclusion, we see that
we have now an alternative way to derive the governing equations of our original \textit{meso} model. In addition to its direct mesoscopic derivation described above in the first paragraph, we can derive it also by constructing first a more microscopic \textit{MESO} model  and then recognizing  the \textit{meso} model as a pattern in  solutions to its governing equations. This new way of deriving the \textit{meso} model  seems to be complicated and indeed, it is  rarely used. Nevertheless, it is important that this alternative way of derivation exists and that, by following it,  we arrive at least at two new results: (a) the material parameters $\zeta$ through which the individual nature of macroscopic systems is expressed in  the \textit{meso} model (\ref{Gdyn}) appear as  functions  of the material parameters playing the same role in the more microscopic \textit{MESO} model, and (b) the reducing dynamics, giving rise to thermodynamics (as we show in Section \ref{RD}).

The above consideration motivates us to start our investigation of externally forced macroscopic systems with two mesoscopic  models instead of with only one such model
(\ref{Gdyn}). The second model (\textit{MESO} model) is formulated on a more microscopic level than the level on which the model (\ref{Gdyn}) is formulated. By "a more microscopic model" we mean that  more details are taken into account in the model.
We write the governing equations of the second model formally as
\begin{equation}\label{Fdyn}
\dot{x}=G(x,\varsigma, \mathcal{F}^{MESO})
\end{equation}
where $x$ denotes state variables, $\varsigma$ material parameters and $\mathcal{F}^{MESO}$ the external influence. The state space used in the \textit{meso} model (\ref{Gdyn}) is denoted by the symbol $N$ (i.e. $y\in N$ ) and the state space used on the more microscopic \textit{MESO} model (\ref{Fdyn}) is denoted by the symbol $M$ (i.e. $x\in M$).  We shall call hereafter  the dynamics described by (\ref{Fdyn}) as \textit{MESO} dynamics and the dynamics described by (\ref{Gdyn}) by \textit{meso} dynamics.

How do we formulate the \textit{MESO} model (\ref{Fdyn})? In its direct derivation we  proceed in the same way as we do in the direct derivation of the {meso} model (\ref{Gdyn}). The difference is only in that the \textit{meso } measurements are replaced by more detailed \textit{MESO} measurements  and that the same type of physics as the one expressed in (\ref{Gdyn}) is now  expressed in (\ref{Fdyn}) in a more detail.

As an example of \textit{meso} dynamics  (\ref{Gdyn})    we can take Boussinesq equations describing, on the level of fluid mechanics (i.e. the \textit{meso  level} in this example is the level of fluid mechanics), the Rayleigh-B\'{e}nard system. The corresponding to it \textit{MESO level} could be the level of kinetic theory on which the state variable $x$ is the one particle distribution function and Eq.(\ref{Fdyn}) is a kinetic equation expressing the same physics as the one expressed in the Boussinesq equations but on the level of kinetic theory.

Having  both \textit{MESO}  and \textit{meso} dynamics,  we are in position to provide a new derivation of the \textit{meso} dynamics (\ref{Gdyn}) and also to identify reducing $MESO\rightarrow meso$ dynamics that, as we shall see below in Section \ref{RDMmde}, provides us with a new \textit{meso thermodynamics}. The process leading from \textit{MESO level} to \textit{meso level} is conveniently seen (see Section \ref{EX2}) as a pattern recognition in the \textit{MESO} phase portrait.  By \textit{MESO} phase portrait  we  mean a collection of trajectories (i.e. solutions to (\ref{Fdyn}) ) passing through all $x\in M$ for a large family of the material parameters $\xi$ and external forces $\mathcal{F}^{MESO}$. The pattern that we search is the one which can be interpreted as representing the mesoscopic phase portrait corresponding to the \textit{meso} dynamics (\ref{Gdyn}).
We prefer to refer to  the process involved in  the passage from \textit{ MESO} to \textit{meso} dynamics as  a pattern recognition process rather than the more frequently used  "coarse graining" process  since the latter term evokes   procedures (as e.g. making pixels and averaging in them) that are manifestly coordinate dependent and thus geometrically (and consequently also physically) meaningless.

\section{Reducing Dynamics, Thermodynamics }\label{RD}

We now proceed to investigate the pattern recognition process leading from \textit{MESO} dynamics to \textit{meso} dynamics.
We recognize  first its  complexity. We recall for instance that this type of investigation  constitutes  in fact   the famous Hilbert's 6th problem (see \cite{GKHilb}).
Roughly speaking, any investigation of the $MESO\,\rightarrow\,meso $ passage consists essentially in splitting the \textit{MESO} dynamics (\ref{Fdyn})  into the \textit{meso} dynamics (\ref{Gdyn}) (that we call  \textit{reduced dynamics} if we regard it in the context of $MESO\,\rightarrow\,meso$ passage) and another dynamics that makes the reduction (that we  call \textit{reducing dynamics}). While  most investigations of  the $MESO\,\rightarrow\,meso $ passages have  focused  in the past on  the reduced dynamics, we  show that investigations of the reducing dynamics are also interesting and bring in fact an additional important information that  can be  interpreted  as  an introduction of thermodynamics on the \textit{meso} level. The reduced dynamics (i.e. \textit{meso} dynamics) together with the thermodynamics  implied by the reducing dynamics  express then (on \textit{meso  level})  the complete physics of the macroscopic system under consideration.

More details of the behavior
of the macroscopic systems under consideration are seen on the \textit{MESO} level (represented by (\ref{Fdyn}) ) than on the \textit{meso} level.
Let $\mathcal{P}^{MESO}$ and $\mathcal{P}^{meso}$ be the phase portraits corresponding to  the \textit{MESO} dynamics (\ref{Fdyn}) and the \textit{meso} dynamics (\ref{Gdyn}) respectively. Our problem is to recognize $\mathcal{P}^{meso}$ as a pattern inside  $\mathcal{P}^{MESO}$.
In the pattern recognition process we recover   the less detailed viewpoint expressed in (\ref{Gdyn}) (that arises in the pattern recognition process as the reduced dynamics) but  in addition we also begin to see the reducing dynamics making the pattern to emerge. In this section we argue that  the reducing dynamics, is in its essence thermodynamics.
In order to be able to justify the use of the term "thermodynamics"   we begin by recalling  the standard (i.e. Gibbs) formulation of the classical thermodynamics and show subsequently  that   the reducing dynamics is indeed its natural extension. The level of description used in the classical equilibrium thermodynamics is called in this paper \textit{equilibrium level}.

In this section we  concentrate on establishing a unified formulation of the reducing dynamics. We show that the formalism puts under a single
umbrella the  thermodynamics of driven systems and well established classical, microscopic, and mesoscopic equilibrium and nonequilibrium
thermodynamics. The unification power of the formalism   is in this section the principal argument  supporting it. In the following section (Section \ref{EX}) we then collect illustrative examples and applications providing additional support.

\subsection{Classical equilibrium thermodynamics; statics}\label{RDET}

The point of departure of the classical equilibrium thermodynamics is the postulate

\textit{\textbf{equilibrium Postulate 0}}

 of the \textit{existence of equilibrium states}. For example,  Callen formulates \cite{Callen} it as follows: [\textit{"... in all
 systems there is a tendency to evolve toward states in which the properties are determined by intrinsic factors and not by previously
 applied external influences. Such simple terminal states are, by definition, time independent. They are called equilibrium states..."}].
 The level of description on which  investigations  are limited only to macroscopic systems  at equilibrium states will be
 called \textit{equilibrium} level. No time evolution takes place on this level.

The next postulate
addresses the \textit{state variables} used on  \textit{equilibrium} level to characterize  the equilibrium states introduced in the previous postulate.

\textit{\textbf{equilibrium Postulate I}}

\textit{The state variables on \textit{equilibrium} level are
the state variables needed to formulate overall macroscopic mechanics (the number of moles $N$,  the volume $V$, and the macroscopic mechanical kinetic energy $E_{mech}$)  and in addition the internal energy $E_{int}$ that is a new, extra mechanical quantity,
serving  as an independent state variable. The internal energy $E_{int}$ then combines with the macroscopic mechanical $E_{mech}$ to define the
overall total  energy   $E=E_{mech}+E_{int}$. We shall denote the state variables of the classical equilibrium thermodynamics by the symbol $\omega$  (i.e. $\omega=(E,N,V)$) an the equilibrium state space $\Omega$ (i.e. $\omega\in \Omega$)}.

The third postulate addresses the way the equilibrium states are reached.

\textit{\textbf{equilibrium Postulate II}}

(i) \textit{The fundamental thermodynamic relation consists of three potentials}
\begin{equation}\label{classftr}
N^{(ee)}(\omega); \, E^{(ee)}(\omega);\,S^{(ee)}(\omega)
\end{equation}
The two potentials, namely the number of moles $N^{(ee)}$ and the energy $N^{(ee)}$ are universal:  $N^{(ee)}=N;\,\,E^{(ee)}=E$. The  entropy $S^{(ee)}(\omega)$ is not universal. It is the
 quantity in which,  on \textit{equilibrium level},  the individual nature of the macroscopic systems under consideration are expressed.   The association between $S^{(ee)}(\omega)$ and the macroscopic systems can be obtained, if we remain inside \textit{equilibrium level}, only by experimental observations (whose results are collected  in the so called thermodynamic tables).
The entropy  $S^{(ee)}(E,V,N)$  is required to satisfy the following three properties.  (i) $S^{(ee)}(E,V,N)$ is a real valued and sufficiently regular function of $z$, (ii) $S^{(ee)}(E,V,N)$ is homogeneous of degree one (i.e. $S^{(ee)}(\lambda E,\lambda V,\lambda N)= \lambda S^{(ee)}(E,V,N)$ which means that the energy, number of moles, volume, and entropy are all extensive variables), and (iii) $S^{(ee)}(E,V,N)$ is a concave function (we exclude from our considerations  in this paper critical states and phase transitions).

(ii) \textit{Equilibrium states are defined as states at which   the entropy $S^{(ee)}(\omega)$  reaches its maximum allowed by constraints (i.e. MaxEnt principle on equilibrium level)}.

Since we consider in this paper thermodynamics associated with passages between two general levels, we need a clear notation. The upper index $(ee)$ in potentials introduced in (\ref{classftr})
means $equilibrium \rightarrow  equilibrium$, i.e.   the passage in which the starting level is \textit{equilibrium} level and the target level is also \textit{equilibrium} level. If the passage that we investigate is
\textit{MICRO} $\rightarrow$ \textit{equilibrium}
(in Section \ref{RDMee} below), we  shall use $(MIe)$, if the passage is \textit{MESO} $\rightarrow$ \textit{equilibrium}
(in Section \ref{RDMMee}), we shall use $(Me)$, and in the investigation of the passage \textit{MESO} $\rightarrow$ \textit{meso}
(in Section \ref{RDMmde}), we shall use $(Mm)$. The first letter in the upper index  denotes always the level on which the quantity is defined and the second letter the level to which the reduction aims or the level from which it is reduced (see (\ref{MIimpl}), or (\ref{Meimpl}) below).

In order to write explicitly the MaxEnt principle, we introduce
\begin{equation}\label{Phiclass}
\Phi^{(ee)}(\omega;T,\mu)=-S^{(ee)}(\omega)+E^{*}E^{(ee)}(\omega)+N^{*}N^{(ee)}(\omega)
\end{equation}
called  a thermodynamic potential on \textit{equilibrium} level. By $\omega^*=(E^*,N^*,V^*)$ we denote conjugate state variables; $E^{*}$  is conjugate to $E$ (i.e. $E^*=S^{(ee)}_E$),  $N^*$ is conjugate to $N$ (i.e. $N^*=S^{(ee)}_N$), and $V^*$ is conjugate to $V$ (i.e.$V^*=S^{(ee)}_V$). We use hereafter the shorthand notation $S_E=\frac{\partial S}{\partial E}$,... .
In the classical equilibrium thermodynamics the conjugate variables  have particular
  names,  namely,  $E^*=\frac{1}{T}, N^*=S_N=-\frac{\mu}{T}, V^*=-\frac{P}{T}$, where $T$ is the temperature, $\mu$ the chemical potential,
  and $P$ the pressure.

Entropy  $S^{(ee)}(E,V,N)$
  transforms, under the  Legendre transformation, into its conjugate  $S^{(ee)*}(\mu,T)$,
\begin{equation}\label{eqimp}
S^{(ee)*}(\mu,T)=[\Phi^{(ee)}(\omega;T,\mu)]_{\omega=\omega_{eq}(T,\mu)}
\end{equation}
where $\omega_{eq}(T,\mu)$ is a solution of $\Phi^{(ee)}_{\omega}=0$.
As a direct consequence of the homogeneity of $S^{(ee)}$,
 $S^{(ee)*}(T,\mu) =-\frac{P}{VT}$.

We note that the MaxEnt principle in the classical equilibrium thermodynamics does not address the time evolution leading to the equilibrium states (i.e. it does not address the process of
preparing macroscopic systems to equilibrium thermodynamic observations). It addresses only the question of what is the final result of such time evolution. We shall introduce such time evolution later in this paper.

\subsection{MICRO $\rightarrow$ equilibrium; Gibbs equilibrium statistical mechanics; statics }\label{RDMee}

Another part of the classical equilibrium theory  is the Gibbs equilibrium statistical mechanics that investigates the passage  $MICRO \rightarrow equilibrium$. We shall formulate the physical basis of the Gibbs theory again in three postulates that are  direct adaptations of  the three postulates in Section \ref{RDET}  to \textit{MICRO level}.

The first postulate, Postulate 0,  is the same as in the classical equilibrium thermodynamics except that we include in it  the statement that  \textit{MICRO level} is also well established.

The second postulate addresses the state variables

\textit{\textbf{MICRO$\rightarrow$ equilibrium Postulate I}}.

\textit{State variables on MICRO level
are position vectors $\rr=(\rr_1,...,\rr_N)$ and momenta $\vv=(\vv_1,...,\vv_N)$ of $N$ particles, $N\sim 10^{23}$, (or alternatively the $N$-particle distribution function $f(\rr,\vv)$)}.

Next, we proceed to the third postulate that addresses the time evolution.  Since the reduced time evolution in the passage \textit{MICRO} $\rightarrow$ \textit{equilibrium} is no time evolution,  the time evolution taking place on  \textit{MICRO level} is the reducing time evolution. The \textit{MICRO level} time evolution $(\rr,\vv)_0\mapsto(\rr,\vv)_t$ is governed by Hamilton's
equations $\left(\begin{array}{cc}\dot{\rr}\\ \dot{\vv}\end{array}\right)=\left(\begin{array}{cc}0&1\\-1&0\end{array}\right)\left(\begin{array}{cc}E^{(MICRO)}_{\rr}\\E^{(MICRO)}_{\vv}\end{array}\right)$, where $E^{(MICRO)}(\rr,\vv)$ is the microscopic energy. This microscopic time evolution induces the time evolution $f_0(\rr,\vv)\mapsto f_t(\rr,\vv)=f_0((\rr,\vv)_{-t})$. In the Gibbs equilibrium statistical mechanics only two
aspect of the \textit{MICRO} time evolution are retained: (1) conservations of the total mass $N^{(MIe)}(f)$ and the total energy $E^{(MIe)}(f)$ defined below in (\ref{microftr}), and (2) an assumption about the \textit{MICRO} trajectories, namely an ergodic-type hypothesis. The second postulate is thus the following.

\textit{\textbf{MICRO $\rightarrow$ equilibrium Postulate II}}

(i) \textit{The fundamental thermodynamic relation consists of three potentials
\begin{eqnarray}\label{microftr}
N^{(MIe)}(f)&=&\int d\rr\int d\vv f(\rr,\vv)\nonumber \\
E^{(MIe)}(f)&=&\int d\rr\int d\vv E^{(MICRO)}(\rr,\vv) f(\rr,\vv)\nonumber \\
S^{(MIe)}(f)&=&-k_B\int d\rr\int d\vv f(\rr,\vv)\ln f(\rr,\vv)
\end{eqnarray}
where  $k_B$ is the Boltzmann constant, $N^{(MIe)}(f)$   has the physical interpretation of number of moles, $E^{(MIe)}(f)$  is the energy.}  The map leading from the state space of the Liouville representation of classical mechanics to the state space of the classical equilibrium thermodynamics will be denoted by the symbol $\mathfrak{P}^{(MIe)}$, i.e.
\begin{equation}\label{PMIe}
f\mapsto \mathfrak{P}^{(MIe)}(f)=(N^{(MIe)}(f),E^{(MIe)}(f))
\end{equation}

(ii)  \textit{$N^{(MIe)}(f)$ and $E^{(MIe)}(f)$ introduced in the fundamental thermodynamic relation (\ref{microftr})  are conserved during the time evolution}.

(iii) \textit{Particle trajectories $(\rr,\vv)_0\mapsto(\rr,\vv)_t$ fill up the microscopic phase space $M^{(MICRO)}$ (i.e. $(\rr,\vv)\in M^{(MICRO)}$)  so that time averages can be replaced with  averages (by using certain measures)  in  $M^{(MICRO)}$ (\textit{the ergodic hypothesis})}.

(iv)\textit{Equilibrium states are defined as states at which   $S^{(MIe)}(f)$  reaches its maximum allowed by constraints (i.e. MaxEnt principle for  the MICRO $\rightarrow$ equilibrium passage). The expression (\ref{microftr}) for $S^{(MIe)}(f)$ is in the Gibbs theory universally valid for all macroscopic systems.   The quantity that on  MICRO level expresses the individual nature of the macroscopic systems under consideration is only the energy $E(f)$}.

In order to write explicitly the MaxEnt principle on the \textit{MICRO} level, we introduce, as we did in the previous section, the thermodynamic potential
\begin{equation}\label{Phi}
\Phi^{(MIe)}(f;T,\mu)=-S^{(MIe)}(f)+\frac{1}{T}E^{(MIe)}(f)-\frac{\mu}{T}N^{(MIe)}(f)
\end{equation}

The fundamental thermodynamic relation on \textit{equilibrium level} implied by the fundamental thermodynamic relation (\ref{microftr}) on \textit{MICRO level} is given by
\begin{eqnarray}\label{MIimpl}
N^{(eMI)}(\omega)&=&[N^{(MIe)}(f)]_{f=f_{eq}}\nonumber \\
E^{(eMI)}(\omega)&=&[E^{(MIe)}(f)]_{f=f_{eq}}\nonumber \\
S^{(eMI)*}(\mu,T)&=&[\Phi^{(MIe)}(f,T.\mu)]_{f=f_{eq}}=-\frac{P}{VT}
\end{eqnarray}
where $f_{eq}(\rr,\vv;T,\mu)$,
solutions of $\Phi^{(MIe)}_{f(\rr,\vv)}=0$,  are equilibrium states.  They form a manifold $\mathcal{M}_{eq}\subset M$ (i.e. $f_{eq}(\rr,\vv;T,\mu)\in \mathcal{M}_{eq}$) that is an invariant manifold with respect to the \textit{MICRO} time evolution. There is no time evolution that  takes place on $\mathcal{M}_{eq}$. The upper index $(eMI)$ means that the quantity belongs to \textit{equilibrium level} and is obtained from an analysis taking place on \textit{MESO level}. This notation was already introduced in the text following Eq.(\ref{classftr}).

We note that the MaxEnt principle on \textit{MICRO level} (i.e. $MICRO  \rightarrow equilibrium$  Postulate II), as well as the equilibrium Postulate II in the classical equilibrium thermodynamics (see Section \ref{RDET}), does not really
address the time evolution leading to equilibrium sates. The \textit{MICRO} time evolution $f_0(\rr,\vv)\mapsto f_t(\rr,\vv)=f_0((\rr,\vv)_{-t})$ introduced in the $MICRO\rightarrow equilibrium$  postulate above leaves the Gibbs entropy introduced in (\ref{microftr}) unchaged (see more in Section \ref{EX2}).
As on  \textit{equilibrium level},  the  $MICRO  \rightarrow equilibrium$  Postulate II addresses only the final result of such evolution. In Section \ref{EX2} we shall
address the $MICRO\rightarrow equilibrium$ reducing time evolution. We shall  write down explicitly the equations governing it.

The Gibbs $MICRO\rightarrow equilibrium$ theory enriches the classical equilibrium thermodynamics in particular in the following two points: (i) it brings
a microscopic insight into the meaning of the internal energy,  and (ii) it offers a way to calculate the fundamental thermodynamic relation from the knowledge of microscopic interactions.

Regarding the first point, we note that in the context of \textit{MICRO level} the internal energy  is the energy of the particles modulo the overall mechanical energy. The mechanical  origin of the
internal energy  implies then  the mechanical nature of the heat and consequently the energy conservation law involved in $equilibrium$ Postulate II.

As for the fundamental thermodynamic relation, the Gibbs equilibrium statistical mechanics (specifically the MaxEnt
Principle in Postulate II of the Gibbs theory)  provides a mapping between the fundamental thermodynamic relation (\ref{microftr}) on \textit{MICRO level} (note  that it is the particle energy $E^{(MICRO)}(\rr,\vv)$
through which the individual nature of macroscopic systems is expressed  on \textit{MICRO level}) and the equilibrium
fundamental thermodynamic relation (\ref{classftr}) (note that  the quantity through which the individual nature of macroscopic systems is expressed on  \textit{equilibrium} level is the entropy $S^{(ee)}(N,V,E))$.

Finally, we note that  the Gibbs equilibrium theory is not  supported by a rigorous analysis of the \textit{MICRO}  mechanics. Both the ergodic-like behavior of particle trajectories and the tendency  of $S^{(MIe)}(f)=-k_B\int d\rr\int d\vv f(\rr,\vv)\ln f(\rr,\vv)$ to reach its maximum (allowed by constraints) during the microscopic time evolution remain, for most macroscopic systems, an unproven assumption.  The  support for the Gibbs theory comes  from plausible assumptions, illustrations,
and the success of its applications. The same will be then true for the general formulation of reducing dynamics presented below.

\subsection{MESO $\rightarrow$ equilibrium; statics }\label{RDMMee}

So far, we have considered only the ultimate microscopic level (called \textit{MICRO level}) and the ultimate macroscopic level (called \textit{equilibrium  level}).
Now we take into consideration also mesoscopic levels  and formulate a general thermodynamics associated with the passage
\textit{MESO} $\rightarrow$ \textit{meso} (that we call hereafter simply thermodynamics).
We begin with Postulate 0. We modify it by noting that
\textit{equilibrium level} is not the only well established level that is less microscopic than the \textit{MICRO level}.  There is in fact a whole family of
such levels (for example fluid mechanics and kinetic theory levels). These well established mesoscopic levels  differ from the equilibrium level
by the fact that, in general, a time evolution  takes place on them (we recall that no time evolution takes place on the equilibrium level) and also
by the fact that they are applicable also to macroscopic systems subjected to external influences (e.g. the level of fluid mechanics is applicable to
the Rayleigh-B\'{e}nard system). We thus replace the postulate of the existence of equilibrium states with a more general

\textit{\textbf{ MESO Postulate 0}}

\textit{ There exist well  established  mesoscopic levels }.

The remaining two postulates are the same as in the Gibbs theory except that the state variable $f(\rr,\vv)$ used on the \textit{MICRO level} is replaced by another state variables used on mesoscopic
levels. As we have done already in Eq.(\ref{Fdyn}), we shall denote it on $MESO$ level by the symbol $x$. For example, $x$ can be one particle distribution function (used in kinetic theory) of
hydrodynamic fields (used in fluid mechanics):

\textit{\textbf{MESO $\rightarrow$ equilibrium Postulate I}}.

\textit{ State variables on MESO level are quantities denoted by the symbol $x$}

We recall the $\textit{MESO level}$ is a well established level (i.e. theoretical predictions on \textit{MESO level} agree with results of \textit{MESO} experimental observations). This then means  that  $x$  is known.

\textit{\textbf{MESO $\rightarrow$ equilibrium Postulate II (statics)}}.

(i) \textit{ The fundamental thermodynamic relation consists of a specification of three potential
\begin{equation}\label{MESOftr}
N^{(Me)}(x),E^{(Me)}(x),S^{(Me)}(x)
\end{equation}
denoting the number of moles, energy, and entropy respectively.} The map leading from \textit{MESO} state space $M$  to the state space of the classical equilibrium thermodynamics will be denoted by the symbol $\mathfrak{P}^{(Me)}$, i.e.
\begin{equation}\label{PMe}
x\mapsto\mathfrak{P}^{(Me)}(x)=(N^{(Me)}(x),E^{(Me)}(x))
\end{equation}
(compare with (\ref{PMIe})).

(ii) \textit{Equilibrium states are defined as states at which   the entropy $S^{(Me)}(x)$
 reaches its maximum allowed by constraints (i.e. MaxEnt principle for  the MESO $\rightarrow$ equilibrium passage)}.

As in previous sections, we introduce   thermodynamic potential
\begin{equation}\label{Phi1}
\Phi^{(Me)}(x;T,\mu)=-S^{(Me)}(x)+\frac{1}{T}E^{(Me)}(x)-\frac{\mu}{T}N^{(Me)}(x)
\end{equation}
Equilibrium state $x_{eq}$ are states at which $\Phi^{(Me)}(x;T,\mu)$ reaches its minimum.
Consequently, $x_{eq}$ are solutions to
\begin{equation}\label{Phieq}
\Phi^{(Me)}_x(x,T,\mu)=0,
\end{equation}
Such states, called equilibrium states,  form  equilibrium a manifold denoted by the symbol $\mathcal{M}_{eq}\subset M$.

The fundamental thermodynamic relation on \textit{equilibrium level} implied by the fundamental thermodynamic relation (\ref{MESOftr}) on \textit{MESO level} is given by
\begin{eqnarray}\label{Meimpl}
N^{(eM)}(\omega)&=&[N^{(Me)}(x)]_{x=x_{eq}}\nonumber \\
E^{(eM)}(\omega)&=&[E^{(Me)}(x)]_{x=x_{eq}}\nonumber \\
S^{(eM)*}(\mu,T)&=&[\Phi^{(Me)}(x,T.\mu)]_{x=x_{eq}}=-\frac{P}{VT}
\end{eqnarray}
where $x_{eq}(\rr,\vv;T,\mu)$, equilibrium states,  are
solutions of (\ref{Phieq}).  The upper index $(eM)$ means that the quantity belongs to \textit{equilibrium level} and is obtained from an analysis taking place on \textit{MESO level}. This notation was already introduced in the text following Eq.(\ref{classftr}).

Summing up, the difference between the Gibbs equilibrium statistical mechanics and the  mesoscopic equilibrium theory formulated above is in the Postulate 0, in the fundamental
thermodynamic relation, and in the arguments supporting the theory. Postulate 0 includes now also existence of mesoscopic levels. Regarding the fundamental thermodynamic relation,
 all three potentials $N^{(Me)}(x),E^{(Me)}(x)$, and $S^{(Me)}(x)$
have to be specified. The same three potentials have to be also specified in the Gibbs theory (see (\ref{microftr})) but
two of them, namely $N^{(MIe)}$ and $S^{(MIe)}$,  are universal.  On \textit{MESO  level},  neither  of them is universally applicable.
For example, let $x$ be one particle distribution function.  The fundamental thermodynamic relation (\ref{microftr}),  but now transposed to the  level of kinetic theory (i.e. the
N-particle distribution function is  replaced by one particle distribution function),  leads to the fundamental thermodynamic relation representing ideal gas on \textit{equilibrium} level
(recall that if
 $f$ in (\ref{microftr}) is replaced by one particle distribution function then the only energy is the kinetic energy); in order to
include more complex macroscopic systems, e.g. van der Waals gas, one has to modify both the energy - by introducing a mean field energy - and entropy - see more in \cite{Gr71} where also the corresponding reducing dynamics is specified). As for the supporting arguments, they now mainly come  from relating  the \textit{MESO} equilibrium theory to the Gibbs theory. \textit{MESO} equilibrium theories
are indeed an organic part of the Gibbs equilibrium statistical mechanics. They arise as   its  simplified versions applicable to particular families of macroscopic systems.

\subsection{MESO $\rightarrow$ equilibrium; reducing dynamics }\label{RDMMde}

An important advantage of investigating the passage $MESO \rightarrow equilibrium $ instead of the passage $MICRO \rightarrow equilibrium $ is that we can more easily investigate
reducing dynamics. We have seen in Section \ref{RDMee} that in order to pass from $MICRO$ dynamics to \textit{equilibrium  level}, we need assumptions (that,  at least in general, remain unproven)
about ergodic-type behavior of microscopic trajectories. On the other hand,  mesoscopic-type experimental observations include also direct observations of the approach to equilibrium. Based on
results of such  observations, mathematical formulations of particular examples of the reducing dynamics $MESO \rightarrow equilibrium $ have been developed (for example the Navier-Stokes-Fourier equations of fluid mechanics and the Boltzmann kinetic equation of gas dynamics). The Boltzmann kinetic equation
 was then the first time evolution equation for which the passage $kinetic\,\,theory \rightarrow equilibrium$ was  explicitly investigated (by Ludwig Boltzmann).
In investigations of the reducing dynamics representing the passage   $MESO \rightarrow equilibrium $ we can therefore  use result obtained  independently in several particular examples of well-established mesoscopic dynamical theories.

The abstract formulation of reducing dynamics $MESO \rightarrow equilibrium$ presented below  has  emerged as a common mathematical
structure of such well established theories.  The first step was made by Clebsch \cite{Clebsch}, who realized that the particle dynamics
and the Euler fluid mechanics share the structure of Hamiltonian dynamics.  The investigation initiated by Clebsch then
continued in particular in the works of Arnold \cite{Arnold}, Marsden and Weinstein \cite{MW}. Independently,
Landau and Ginzburg \cite{LG} and Cahn and
Hilliard \cite{CH}  have recognized a common structure of gradient dynamics in the part of the time
evolution that is represented in (\ref{GENERIC})  by the second term on its right-hand side.   Time evolution equations
involving both the Hamiltonian and the
gradient part had appeared first in \cite{DV}, in \cite{Grmboulder} (that was presented at the AMS-IMS-SIAM Joint
Summer Research Conference in the Mathematical Sciences on Fluids and Plasmas:Geometry and
Dynamics, held at the University of Colorado, Boulder, CO, USA, 17–23 July 1983) and in \cite{Morr}, \cite{Kauf}, \cite{GrPhysD},  \cite{BEd}. In
\cite{GO}, \cite{OG} the abstract equation (\ref{GENERIC}) has been called GENERIC. Its  formulation in the context of the contact geometry is
presented in \cite{Grmcontact}
Specific realizations of (\ref{GENERIC}) on many examples of \textit{MESO} levels can be found in \cite{Obook}, \cite{Grmadv}.

Now we proceed to the formulation.
Postulate 0 and Postulate I remain the same as in Section \ref{RDMMee}. In Postulate II we replace the static MaxEnt principle with
dynamics MaxEnt principle. We explicitly specify the dynamics making the maximization of entropy.

The  third postulate \textit{MESO $\rightarrow$ equilibrium Postulate II (statics)} in Section \ref{RDMMee} is now replaced  with dynamic postulate

\textit{\textbf{ MESO $\rightarrow$ equilibrium Postulate II (dynamics)}}

(i) \textit{  remains the same as in MESO $\rightarrow$ equilibrium Postulate II (statics)}

(ii) \textit{The time evolution making the passage  MESO $\rightarrow$ equilibrium  is governed by the GENERIC equation}
\begin{equation}\label{GENERIC}
\dot{x}=[TL(x)x^*-[\Xi_{x^*}(x,X^{(CR)}(x^*))]_{x^*=\Phi^{(Me)}_x}
\end{equation}

In the rest  of this section we explain the meaning of the symbols appearing in (\ref{GENERIC}) and  prove that the time evolution
governed by (\ref{GENERIC}) brings $x$ indeed  to the equilibrium states $x_{eq}\in\mathcal{M}_{eq}$ that are solutions of $\Phi^{(Me)}_x=0$.

The first term on the right hand side of (\ref{GENERIC}) represents the part of the time evolution of $x$ that is on \textit{MESO level} directly
inherited from \textit{MICRO level}. It generates the Hamiltonian time evolution. The operator $L$, transforming the covector $x^*$ into a vector,
is a Poisson bivector. This means that $\{A,B\}=<A_x,LB_x>$ is a Poisson bracket, i.e.  $\{A,B\}=-\{B,A\}$ and the Jacobi
identity $\{A,\{B,C\}\}+\{B,\{C,A\}\}+\{C,\{A,B\}\}=0$ holds. By the symbols $A,B,C$ we denote sufficiently regular real
valued functions of $x$, $<,>$ denotes pairing in $M$. We recall that on \textit{MICRO level} the Hamiltonian time evolution is generated by energy $E(x)$.
We recall that in the particular case if $x=(r,v)^T$, where $r$ is the particle position vector, $v$ particle momentum, and $()^T$ means
transpose of  $()$, then $L=\left(\begin{array}{cc}0&1\\-1&0\end{array}\right)$, and the equation governing the time evolution of
$(r,v)$ is $\left(\begin{array}{cc}\dot{r}\\ \dot{v}\end{array}\right)=\left(\begin{array}{cc}0&1\\-1&0\end{array}\right)
\left(\begin{array}{cc}E_r\\ E_v\end{array}\right)$.
In order to keep the energy as a sole generator of dynamics  also on \textit{MESO level}, we require that the operator $L^{(Me)}$ is degenerate
in the sense that
$\{A,S^{(Me)}\}=\{A,N^{(Me)}\}=0$ for all $A$. By using the terminology established in investigations of Hamiltonian systems, the potentials $S^{(Me)}$ and $N^{(Me)}$ are
required to be Casimirs of the Poisson bracket $\{A,B\}$. If this is the case then indeed, the first term on the right hand side of
(\ref{GENERIC}) is $LE^{(Me)}_x$.
A direct consequence of the antisymmetry and the degeneracy of $L$ is that this Hamiltonian part of
the time evolution leaves the energy and the generating potential $\Phi^{(Me)}$ (see (\ref{Phi1})) unchanged (i.e. $(\dot{\Phi}^{(Me)})_{Hamilton}=<\Phi^{(Me)}_x,L\Phi^{(Me)}_x>=<E^{(Me)}_x,LE^{(Me)}_x>=0$).
Examples of the operator $L$ in many \textit{MESO levels} can be
found in \cite{Grmadv}.

Before leaving the Hamiltonian part of the time evolution we make a comment about the role that  the Jacobi identity plays in it. If the
second term on the right hand side of  (\ref{GENERIC}) is absent then the equation $\dot{x}=LE^{(Me)}_x$ governing the time evolution can also be written in
the form $\dot{A}=\{A,E^{(Me)}\}$ holds for all $A$. If we now replace $A$ with $\{A,B\}$ we obtain $\dot{\{A,B\}}=\{\{A,B\},E^{(Me)}\}$. But  $\dot{\{A,B\}}=
\{\dot{A},B\}+\{A,\dot{B}\}=\{\{A,E^{(Me)}\},B\}+\{A,\{B,E^{(Me)}\}\}$. The Jacobi identity guarantees that these two time derivatives of $\{A,B\}$ are equal and thus
the Poisson bracket remains unchanged during the time evolution. If we now consider the time evolution governing by (\ref{GENERIC}) involving
also the second term on the right hand side, the Poisson bracket is not preserved during the time evolution even if the Jacobi identity holds.
The role of the Jacobi identity in GENERIC time evolution is thus much less important that in the Hamiltonian time evolution. Hereafter, we shall
call a time evolution GENERIC time evolution even if the Jacobi identity remains unproven.

The second part on the right hand side of (\ref{GENERIC}) is the part that arises due to the fact that \textit{MESO level} is not \textit{MICRO level}.
This means that  some microscopic details that are seen on \textit{MICRO level} are  ignored on \textit{MESO level}.  This ignorance then influences the time
evolution in such a way that the potential $\Phi^{(Me)}$ approaches its minimum. By $\Xi(x,X)$, called a dissipation potential,  we denote a sufficiently
regular and real valued function of $x\in M$ and of $X$ that is called a thermodynamic force. Its specification $X=X^{(CR)}(x^*)$ as a
function of $x^*$ is called a \textit{constitutive relation}. The superscript "CR" means Constitutive Relation. We assume that the
dissipation potential satisfies the following properties:
\begin{eqnarray}\label{Xi}
&&\Xi(x,0)=0\nonumber \\
&& \Xi\,\,reaches\,\,its\,\,minimum\,\,at\,\,X=0\nonumber\\
&&\Xi\,\,is\,\,a\,\,convex\,\,function\,\,in\,\,a\,\,neighborhood\,\,of\,\,X=0
\end{eqnarray}
Regarding the constitutive relations, we assume that
\begin{equation}\label{crelprop}
<x^*,\Xi_{x^*}(x,X^{(CR)}(x^*))>=\alpha <X^{(CR)},\Xi_{X^{(CR)}}(x,X^{(CR)})>
\end{equation}
where $\alpha >0$ is a parameter.
In addition, we require that the dissipation potential $\Xi$ is degenerate in the following sense:
\begin{eqnarray}\label{degXi}
<[x^*]_{x^*=E^{(Me)}_x},\Xi_{x^*}>=<[x^*]_{x^*=N^{(Me)}_x},\Xi_{x^*}>&=&0\nonumber \\
<x^*,[\Xi_{x^*}]_{x^*=E^{(Me)}_x}>=<x^*,[\Xi_{x^*}]_{x^*=N^{(Me)}_x}>&=&0
\end{eqnarray}

The simplest example of the dissipation potential $\Xi$ satisfying (\ref{Xi}) is the quadratic
potential $\Xi=<X\Lambda X>$, where $\Lambda $ is a matrix with required degeneracy and positive definite if applied on
vectors outside its nullspace. More general potentials arise in particular in chemical kinetics
(see \cite{Grmchem}). It has been suggested in \cite{Beretta} to regard $\Lambda$ as a metric tensor. This interpretation brings then Riemannian geometry to dissipative dynamics. We emphasize that this geometrical viewpoint is limited to the quadratic dissipation potential. In the case of nonlinear dissipation potentials (for example those arising in chemical kinetics - see also Section \ref{EX2}), the geometrical interpretation is still possible but the classical Riemannian geometry has to be replaced by a more general geometry.

 As an example of the constitutive relation satisfying (\ref{crelprop}) we mention the Fourier constitutive relation in
the investigation of heat transfer. In this example $x^*=\frac{1}{T(\rr)}$, where $T(\rr)$ is the local temperature and the constitutive
relation is $X^{(CR)}(x^*)=\nabla \left(\frac{1}{T(\rr)}\right)$. Direct calculations lead to
$<x^*,\Xi_{x^*}>\\=<\frac{1}{T(\rr)},\Xi_{\frac{1}{T(\rr)}}>=\int d\rr\frac{1}{T(\rr)},\Xi_{\frac{1}{T(\rr)}}
=-\int d\rr\frac{1}{T(\rr)},\nabla\Xi_{\nabla\left(\frac{1}{T(\rr)}\right)}\\=
\int d\rr\nabla\left(\frac{1}{T(\rr)}\right),\Xi_{\nabla\left(\frac{1}{T(\rr)}\right)}=<X^{(CR)},\Xi_{X^{(CR)}}>$ provided
the boundary conditions guarantee that the integrals over the boundary that arise in  by parts integrations
(leading to the last equality) equal zero. We see that in this example $\alpha=1$. In the context of chemical kinetics,
where the thermodynamic forces $X$ are chemical affinities, the parameter $\alpha\neq 1$   (see Section \ref{EX1} and \cite{Grmchem};
for example, for the dissipation potential (\ref{XiN}) the coefficient $\alpha=\frac{1}{2}$ and for the dissipation potential $\Xi$
appearing in the Boltzmann equation (\ref{intlin}) the coefficient $\alpha=\frac{1}{4}$).

Dissipation  potentials and  constitutive relations will play an important role also in the investigation of the passage
\textit{MESO} $\rightarrow$ \textit{meso}
in Section \ref{RDMmde} below.

It follows directly from (\ref{GENERIC}) and from the properties of $L$, $\Xi$, and $X^{CR}$  listed above that
\begin{equation}\label{asGEN}
\dot{\Phi}^{(Me)}=-<x^*,\Xi_{x^*}(x,X^{CR})>=-\alpha <X^{CR},\Xi_{X^{CR}}(x,X^{CR})>\leq 0
\end{equation}
The first equality is the required property (\ref{crelprop}) of constitutive relations and the last inequality is a direct consequence of
the properties (\ref{Xi}).
The inequality (\ref{asGEN})  allows us to see the thermodynamic potential $\Phi^{(Me)}$ as a Lyapunov function associated with the approach of
 solutions of (\ref{GENERIC}) to $x_{eq}$ given by (\ref{Phieq}). We have thus proven that Eq.(\ref{GENERIC}) indeed makes the passage
 \textit{MESO} $\rightarrow$ \textit{equilibrium}.

If, in addition, we assume that $L$ and $\Xi$ are degenerate in the sense that $\{S,A\}=0$ for all $A$ (i.e., if we use  the terminology of
Hamiltonian dynamics, the entropy $S^{(Me)}$ is Casimir of the Poisson bracket $\{A,B\}$), and $<E_x,\Xi_{S^{(Me)}_x}>=0$, $<N^{(Me)}_x,\Xi_{S^{(Me)}_x}>=0$
and $<x^*,\Xi_{E^{(Me)}_x}>=0$, $<x^*,\Xi_{N^{(Me)}_x}>=0 \,\,\forall x^*$ then, in addition to the inequality (\ref{asGEN}),  also the following equalities (conservation laws)
\begin{eqnarray}\label{consEGEN}
\dot{E}^{(Me)}(x)&=&0\\\label{consNGEN}
\dot{N}^{(Me)}(x)&=&0
\end{eqnarray}
hold.

\subsection{MESO $\rightarrow$ meso; reducing dynamics }\label{RDMmde}

In this section we come to the main subject of this paper. We consider externally driven macroscopic systems whose time evolution is
governed on \textit{MESO level} by (\ref{Fdyn}). External forces prevent approach to equilibrium which means that  \textit{equilibrium level}  is
inaccessible and the approach $MESO \rightarrow equilibrium$ does not exist. Let  however  the behavior of the
 externally driven macroscopic
systems under consideration be  found to be  described well also on $meso$ level that is more macroscopic (i.e. it takes into account
less details) than \textit{MESO level}. This then means that by investigating solutions of the governing equations on \textit{MESO level} we have to be
able to recover the governing equations on \textit{meso level}. In addition, such investigation  will also reveal  reducing dynamics making the
passage $MESO \rightarrow meso$. In Section \ref{RDMMde} we have  shown how thermodynamics on  \textit{equilibrium level} arises from the reducing
dynamics $MESO \rightarrow equilibrium$ or $meso \rightarrow equilibrium$. In this section we show how thermodynamics on  \textit{meso level} (we shall call it Constitutive
Relation \textit{meso} thermodynamics or in short form
\textit{\textbf{CR meso-thermodynamics}} to distinguish it from $equilibrium\,\, thermodynamics$ discussed above in Sections \ref{RDET} - \ref{RDMMde})
 arises from the reducing dynamics $MESO \rightarrow  meso$.

We recall that the  formulation of thermodynamics  presented in Section \ref{RDMMde}  (i.e. the formulation of thermodynamics  implied by the reducing
dynamics  $MESO \rightarrow equilibrium$)  has been   supported mainly by the unification that it brings to various versions of mesoscopic
thermodynamics that have emerged in the last one hundred fifty years in  well studied and essentially independently developed (each on the basis if
its own experimental evidence)  mesoscopic dynamical theories. We do not find  such examples in the context of the
passage $MESO \rightarrow meso$. We do find however important results in nonequilibrium thermodynamics, like for instance dissipation thermodynamics
(see references in \cite{GPK}) and extended  thermodynamics (see e.g. \cite{Joubook}, \cite{MullRugg}). We expect them to become, in some form,
 a part of the general formulation of \textit{CR meso-thermodynamics}. Our goal is thus to formulate  \textit{CR meso-thermodynamic} in such a way that  equilibrium thermodynamics,
 dissipation thermodynamics, and extended  thermodynamics make appearance as   its different aspects.

\subsubsection{Motivating example}\label{motex}

Before formulating the three postulates of $MESO\rightarrow meso$ thermodynamics, we work out a particular example.
First, we present the  physical idea and then we formulate it mathematically .
We begin with a given \textit{meso level} represented by   (\ref{Gdyn}). For example, we can think of Eq.(\ref{Gdyn}) as standing  for
the Navier-Stokes-Fourier set of equations.   The corresponding to it more detailed \textit{MESO level} represented by Eq.(\ref{Fdyn}) will
be constructed as an extension of (\ref{Gdyn}).
Following  \cite{MullRugg}, \cite{Joubook},  the extension from \textit{meso} to \textit{MESO} is made, roughly speaking,  by replacing
the second term on the right
hand side of (\ref{GENERIC}) (i.e. the dissipative term)  with   a new state variable (denoted by the symbol $J$ and interpreted physically as a flux corresponding to the state variable $y$). The time evolution of $J$ is then
governed by a newly introduced  equation that involves dissipative term and is coupled to the time evolution of $y$. We require that $J$
dissipates rapidly to a quasi-stationary
state $z_{qeq}(y)$
at  which it becomes completely enslaved to $y$. At such quasi-stationary state, the newly constructed \textit{MESO} dynamics reduces to  the original
\textit{meso} dynamics represented by (\ref{Gdyn}). We shall make now an additional requirement.
Having realized that all   equations governing the reducing time evolution $MESO \rightarrow equilibrium$ possess the structure (\ref{GENERIC}),
 we require that
in the absence
of the dissipative term the  time evolution of $(y,J)$ is Hamiltonian.

In this example we restrict ourselves  to  \textit{meso} dynamics (\ref{Gdyn}) that is GENERIC  (\ref{GENERIC}) and  without
the Hamiltonian part. Moreover, we consider only isothermal systems (see also Section \ref{EX3}) and, for the sake of simplicity,  we omit
the potential $N$ representing the number of moles. The time evolution equation (\ref{Gdyn})  takes thus the form
\begin{equation}\label{I0}
\dot{y}=-[\Xi^{(me)}_{y^*}(y,y^*)]_{y^*=\Phi^{(me)}_y}
\end{equation}
where $\Phi^{(me)}(y,T)$ respectively $\Xi^{(me)}(y,y^*)$ is the thermodynamic potential respectively the dissipation potential
 associated with \textit{meso} $\rightarrow$ \textit{equilibrium} passage.
The temperature $T$ is a constant.

We investigate first the  passage \textit{meso} $\rightarrow$ \textit{equilibrium}.
We  see immediately that (\ref{I0}) implies $\dot{\Phi}^{(me)}=-\left[y^*\Xi^{(me)}_{y^*}\right]_{y^*=\Phi^{(me)}_y}\leq 0$ provided $\Xi^{(me)}$
satisfies the properties (\ref{Xi}). This thermodynamic potential then implies the fundamental thermodynamic relation on \textit{equilibrium level}
\begin{equation}\label{I20}
\Phi^{(em)*}(T)= [\Phi^{(me)}(y,T)]_{y=y_{eq}(T)}
\end{equation}
where $y_{eq}(T)$ is a solution of $\Phi^{(me)}_y=0$. By the upper index $(em)$ in $\Phi^{(em)*}(T)$ appearing in  (\ref{I20}) we denote (see the paragraph following Eq.(\ref{classftr})) that this quantity belongs to \textit{equilibrium level} and is obtained by MaxEnt reduction from \textit{meso level}.

So far, we have looked from \textit{meso level} to \textit{equilibrium level}.
Now we   look in   the opposite
 direction towards  \textit{MESO level} involving more details.
We extend the \textit{meso} dynamics (\ref{I0}) to \textit{MESO} dynamics by
following the physical considerations sketched in the beginning of this section. The state variables $x$ on \textit{MESO level} become  $x=(y,J)$,
where $J$ is a newly adopted state variable having the physical interpretation of a "flux" of $y$. Equation (\ref{Fdyn}) is proposed to have the form
\begin{eqnarray}\label{I3}
\dot{y}&=& \Gamma [J^*]_{J^*=\Phi^{(Me)}_J}\nonumber \\
\dot{J}&=& -\Gamma^T [y^*]_{y^*=\Phi^{(Me)}_y} - [\Theta^{(Me)}_{J^*}(y,J^*)]_{y^*=\Phi^{(Me)}_y;J^*=\Phi^{(Me)}_J}
\end{eqnarray}
where $\Gamma$ is an operator, $\Gamma^T$ is its transpose, $\Phi^{(Me)}(y,J)$ is the thermodynamic potential associated with the
\textit{MESO} $\rightarrow$ \textit{equilibrium} passage. The dissipation potential $\Theta^{(Me)}(y,J^*)$ is the Legendre transformation of the dissipation potential $\Xi^{(Me)}(y,X^*)$ where
$X^*=\Theta^{(Me)}_{J^*}$ (i.e. $\Theta^{(Me)}(y,J^*) =[-\Xi^{(Me)}(y,X^*) +X^*J^*]_{X^*=X_0^*(y,J^*)}$, where $X_0^*(y,J^*)$ is a solution of $[-\Xi^{(Me)}(y,X^*) +X^*J^*]_{X^*}=0$).
If  $\Xi^{(Me)}(y,X^*)$ satisfies the properties (\ref{Xi}) then also $\Theta^{(Me)}(y,J^*)$ satisfies them and vice versa.

The time evolution  equation (\ref{I3}) is again GENERIC (\ref{GENERIC}) but contrary to (\ref{I0})  it has now also
the Hamiltonian part \\ $\left(\begin{array}{cc}0&\Gamma\\-\Gamma^T&0\end{array}\right)\left(\begin{array}{cc}y^*\\J^*\end{array}\right)$. The operator
$\left(\begin{array}{cc}0&\Gamma\\-\Gamma^T&0\end{array}\right)$ is skew symmetric for any operator $\Gamma$ but the corresponding to it bracket
does not necessarily satisfy the Jacobi identity for any $\Gamma$. In view of the remark that we made in Section \ref{RDMMde} about the role of the
Jacobi
identity in GENERIC, we still consider (\ref{I3}) as being GENERIC.

At this point we note that the extension that we made above differs from extensions made in \cite{MullRugg}, \cite{Joubook} by requiring  that
the nondissipative part of the extended equation is Hamiltonian. As a consequence, the flux appearing on
the right hand side of the first equation in (\ref{I3}) is conjugate to the flux appearing on the left hand side of the second equation of (\ref{I3}).
The fact that this feature of the extension is not seen in Refs.\cite{MullRugg} and \cite{Joubook} is that  the
master structure for extensions is in Refs.\cite{MullRugg} and \cite{Joubook} the classical Grad-hierarchy reformulation of the Boltzmann equation
 that addresses only a very special physical
 system (namely the ideal gas) and thus, in terms of our formulation,  only a very special class of functions
 $\Phi^{(Me)}(y,J)$ and $\Phi^{(me)}(y)$.

First, we again establish the passage \textit{MESO} $\rightarrow$ \textit{equilibrium}.
It directly follows
from (\ref{I3}) that
\begin{equation}\label{eprodM}
\dot{\Phi}^{(Me)}=-\left[J^*\Theta^{(Me)}_{J^*}\right]_{J^*=\Phi^{(Me)}_J}\leq 0
\end{equation}
provided $\Theta^{(Me)}$ satisfies the properties (\ref{Xi}).
In the same way as on \textit{meso} level we arrive at
 the fundamental thermodynamic relation on \textit{equilibrium} level
\begin{equation}\label{I21}
\Phi^{(eM)*}(T)=[\Phi^{(Me)}(y,J,T)]_{y=y_{eq}(T); J=J_{eq}(T)}
\end{equation}
where $y_{eq}(T)$ and  $J_{eq}(T)$ are solutions to $\Phi^{(Me)}_y=0$ and $\Phi^{(Me)}_J=0$.
The thermodynamic potential  $\Phi^{(Me)}(y,J)$  represents
  a more detailed picture of the physics taking place
in the macroscopic system under consideration than the picture represented by  $\Phi^{(me)}(y)$.  Depending on the particular forms of
$\Phi^{(Me)}(y,J)$ and $\Phi^{(me)}(y)$, some of the details taken into consideration on \textit{MESO level} may or may not show up in the equilibrium fundamental
thermodynamic relation on \textit{equilibrium level}. In general, the
\textit{equilibrium level} fundamental thermodynamic relations (\ref{I20}) and (\ref{I21}) are not identical.

Next, we reduce (\ref{I3}) to (\ref{I0}). Let the operator $\Gamma$, the dissipation potential $\Theta^{(Me)}$, and the thermodynamic potential
$\Phi^{(Me)}$
be such that $J$
evolves in time
 more rapidly than $y$. If this is the case then we  regard the time evolution governed by (\ref{I3}) as proceeding in two stages. In the first stage
 (the reducing  evolution), the time evolution of $J$ is governed by the second equation in (\ref{I3}) in which $y$ (and thus also $y^*$) are fixed. This
 reducing (fast) time evolution is thus governed by
\begin{equation}\label{I4}
\dot{J}=-\Phi^{(Mm)}_{J^*}
\end{equation}
where
\begin{equation}\label{I5}
\Phi^{(Mm)}(X^{(CR)*}(y^*),J^*)=\Theta^{(Me)}(J^*)-X^{(CR)*}(y^*)J^*
\end{equation}
with the constitutive relation
\begin{equation}\label{CR1}
X^{(CR)*}(y^*)=-\Gamma^T y^*
\end{equation}
In order to distinguish the conjugates with respect to the entropy $\Phi^{(Me)}$ (i.e. $y^*=\Phi^{(Me)}_y; J^*=\Phi^{(Me)}_J$) from conjugates with respect to
the dissipation potential $\Theta$, we do not use the upper index star to denote $\Theta_{J^*}$   but we use, following
the traditional notation
 established in nonequilibrium thermodynamics,
$X^*=\Theta_{J^*}$. Still following the traditional terminology of nonequilibrium thermodynamics, we call $X^*$ the thermodynamic force
corresponding to the thermodynamic flux $J^*$.

Now we turn our attention to solutions of (\ref{I4}). We see immediately that
\begin{equation}\label{I100}
\dot{\Phi}^{(Mm)}=-\Phi^{(Me)}_{JJ}(\Phi^{(Mm)}_{J^*})^2\leq 0
\end{equation}
which means that  $J$ tends, as $t\rightarrow \infty$, to $J_{qeq}^*(y^*)$ that is a solution of
\begin{equation}\label{I101}
\Theta^{(Me)}_{J^*}(J^*)=X^{(CR)*}(y^*)
\end{equation}
We see that the potential $\Theta^{(Me)}$  plays different roles in the analysis of $MESO \rightarrow meso$ and in the analysis of $MESO\rightarrow equilibrium$. In the former analysis it plays  the same role as the thermodynamic potential  $\Phi^{(Me)}$ plays in the investigation of the approach
$MESO \rightarrow equilibrium$ governed by (\ref{I3}). In the latter analysis it  plays the role that  is closely related to the
entropy production (see (\ref{eprodM})).

The relation
\begin{equation}\label{Mmth}
\Phi^{(mM)*}(y^*)=\Phi^{(Mm)}(X^{(CR)*}(y^*),J_{qeq}^*(y^*))=-\Xi^{(Me)}(y,X^{(CR)*}(y^*))
\end{equation}
is  the fundamental thermodynamic relation on \textit{meso level} implied by the fast time evolution governed by (\ref{I4}).

If we insert $J_{qeq}^*$ (i.e. solution of (\ref{I101})) into the first equation in (\ref{I3}) we arrive at
\begin{equation}\label{Xxrelation}
\Xi^{(me)}(y,y^*)=[\Xi^{(Me)}(y,X)]_{X=X^{(CR)*}(y,y^*)}
\end{equation}

The analysis presented above   can be summed up in two results.

\textit{Result 1}

Equation (\ref{I4}) governs the reducing time evolution  (i.e. the time evolution making the reduction $MESO\rightarrow meso$) and (\ref{Mmth}) is the fundamental thermodynamic relation  on \textit{meso} level implied by it.

\textit{Result 2}

The reducing time evolution equation (\ref{I4}) is explicitly related to the \textit{MESO} time evolution equation (\ref{I3}) and to the \textit{meso} time evolution equation (\ref{I0}). The \textit{MESO} dynamics (\ref{I3}) is split into (fast) reducing dynamics (\ref{I4}) followed by (slow) reduced dynamics (\ref{I0}).

\subsubsection{General formulation}\label{GF}

In this section we  formulate Result 1  in a more general context. Result 2 requires a detail specification of \textit{MESO} dynamics and a detail analysis of the phase portrait $\mathcal{P}^{MESO}$ that it generates. Except for  a few simple illustrations  presented in Section \ref{EX}, we shall not attempt in this paper to formulate Result 2 in general terms.

Our objective is to adapt  the three Postulates of  $MESO\rightarrow equilibrium$ thermodynamics (formulated in Sections \ref{RDET}, \ref{RDMee}, \ref{RDMMee}, \ref{RDMMde} above)  to $MESO\rightarrow meso$ thermodynamics. First we note an important difference between the reductions $MESO\rightarrow equilibrium$ and $MESO\rightarrow meso$.
In the former reduction the target level is \textit{equilibrium level}, i.e. a level of description on which no time evolution takes place. In such reducing dynamics, the fundamental thermodynamic relations consist of equilibrium state variables $\omega$ expressed in terms of the state variables used on the initial level and  the entropy driving the reduction (see (\ref{microftr}) and (\ref{MESOftr})). In the latter reduction the target level is \textit{meso level} on which the time evolution does take place. The fundamental thermodynamic relation corresponding to $MESO\rightarrow meso$ reduction must again include the state variables $y$ expressed in terms of $x$ but it must also include the vector field $g$ on \textit{meso level} (see (\ref{Gdyn}) ) expressed in terms of $x$. We present now a setting in which we subsequently formulate the fundamental thermodynamic relation of $MESO \rightarrow meso$ thermodynamics.

We begin with \textit{MESO} dynamics (\ref{Fdyn}) and with the map
\begin{equation}\label{PMm}
\mathfrak{P}^{(Mm)}:M\rightarrow N; x\mapsto y=y(x)
\end{equation}
allowing to express the state variables on \textit{meso level} in terms of the state variables on \textit{MESO level} (compare with (\ref{PMIe})  and  (\ref{PMe}). We apply the map $\mathfrak{P}^{(Mm)}$ on (\ref{Fdyn}) and obtain
\begin{equation}\label{Gdynn}
\dot{y}=\mathfrak{P}(G(x))
\end{equation}
Hereafter, we shall write the right hand side of (\ref{Gdynn}) in the form
\begin{equation}\label{J}
\mathfrak{P}(G(x))=\Gamma(\JJ(x))
\end{equation}
where $\Gamma$ is a fixed  operator and $\JJ(x)=(J_1(x),...,J_n(x))$ are quantities called \textit{\textbf{thermodynamic fluxes}}. For example, if (\ref{Fdyn}) is the Boltzmann kinetic equation (with the one particle distribution function $f(\rr,\vv)$, where $\rr$ is the position vector and $\vv$ momentum of one particle) and the the map $\mathfrak{P}^{(Mm)}$ is a projection on the first five moments in $\vv$ then $\Gamma =-\frac{\partial}{\partial \rr}$ and $\JJ$ are higher order moments. In chemical kinetics (see \cite{Grmchem}) $\Gamma$ is the stoichiometric matrix.

With  (\ref{J}), the time evolution equation (\ref{Gdynn}) takes  the form
\begin{equation}\label{JJJ}
\dot{y}=\Gamma(\JJ(x))
\end{equation}
Its right hand side  does  not represent, at least in general, a vector field on $N$.
In order to be that, it has to be closed, i.e. evaluated at $x=c_{cl}(y)$
\begin{equation}\label{Gdynnc}
\dot{y}=[\Gamma(\JJ(x))]_{x=\hat{x}(y)}
\end{equation}
The map  $y\mapsto \hat{x}(y)$ is a map $N\rightarrow M$ called a \textit{\textbf{closure map}}.
In the formulation of the $MESO \rightarrow meso$ thermodynamics we shall limit ourselves to the \textit{meso} dynamics (\ref{Gdyn}) that has the form (\ref{Gdynnc}) with no restriction on the  closure map. It will be the reducing dynamics that will determine it.

We are now in position to formulate three Postulates of $MESO\rightarrow meso$ thermodynamics.
We shall follow  closely the formulation of the reducing dynamics \textit{MESO}  $\rightarrow$ \textit{equilibrium}  presented in Section \ref{RDMMde}.

\textbf{\textit{MESO $\rightarrow$ meso Postulate 0}}

is the same as the \textit{MESO} $\rightarrow$ \textit{meso}  Postulate 0 in Section \ref{RDMMde}.

\textbf{\textit{MESO $\rightarrow$ meso  Postulate I}}

\textit{ $x\in M$ are state variables on MESO level, $y\in N$ are state variables on meso level and (\ref{Gdynnc}) with unrestricted closure map is a family of the time evolution equations on meso level.}

\textbf{\textit{MESO $\rightarrow$ meso Postulate II}}

(i) \textit{The fundamental thermodynamic relation consists of the specification of the following quantities}
\begin{equation}\label{Mmcr}
y(x),\JJ(x),\Phi^{(0Mm)}(x)
\end{equation}

The first quantity in (\ref{Mmcr}) is the map  $\mathfrak{P}^{(Mm)}:M\rightarrow N$  which expresses the state variables $y$ used on \textit{meso level} in terms of state variables $x$ used on the more microscopic \textit{MESO level}. The reducing time evolution introduced below leaves the space $N$ unchanged.
The quantities $\JJ=(J_1,...,J_n)$ are the  thermodynamic fluxes appearing in the target \textit{meso} dynamics (\ref{Gdynnc}).
The final quantity $\Phi^{(0Mm)}$ is the thermodynamic potential  $\Phi^{(0Mm)}: M\rightarrow \mathbb{R}$. Following  (\ref{Phi}) and (\ref{Phi1}), we  write it in the form
\begin{equation}\label{Phi0}
\Phi^{(0Mm)}(x,\theta)=-S^{(Mm)}(x)+\frac{1}{\theta}W^{(Mm)}(x)
\end{equation}
The motivating example discussed above in Section \ref{motex} and the reducing dynamics discussed below indicate that $[S^{(Mm)}(x)]_{x_{qeq}}$  (where $x_{qeq}$ is  $t\rightarrow \infty$ solution of the reducing dynamics)  has the physical interpretation of the entropy production on  \textit{meso level} and $[W^{(Mm)}(x)]_{x_{qeq}}$ has the physical interpretation of the work per unit time performed by external forces. The quantity $\theta$ is a temperature or a quantity having  the physical dimension of the temperature.

(ii) \textit{The  $MESO\rightarrow meso$ reducing time evolution is governed by}
\begin{equation}\label{CRGENERIC}
\dot{x}=[\theta L^{(Mm)}(x)x^*-[\Xi^{(Mm)}_{x^*}(x,x^*)]_{x^*=\Phi^{(Mm)}_x}
\end{equation}
where
\begin{equation}\label{PhiMm}
\Phi^{(Mm)}(x,\XX)=\Phi^{(0Mm)}(x,\theta)+\sum_{i=1}^{n}X_iJ_i(x)
\end{equation}
is the thermodynamic potential. The quantities $\XX=(X_1,...,X_n)$ are called thermodynamic forces corresponding to the thermodynamic fluxes $\JJ=(J_1,...,J_n)$. As in (\ref{GENERIC}), the operator $L^{(Mm)}$ is a Poisson bivector and the $\Xi^{(Mm)}$ is a dissipation potential. Both $L^{(Mm)}$  and $\Xi^{(Mm)}$  are required to be degenerate so that the space $N$ remains invariant  under the time evolution governed by (\ref{CRGENERIC}).

The same considerations as the ones that led us in Section \ref{RDMMde}) to the conclusion that solutions to (\ref{GENERIC}) have the property $x\rightarrow x_{eq}$ as $t\rightarrow \infty$, lead us the the conclusion that solutions of (\ref{CRGENERIC}) have the property $x\rightarrow x_{qeq}$ as $t\rightarrow \infty$, where $x_{qeq}$ is a solution to
\begin{equation}\label{qeq}
\Phi^{Mm)}_x=0
\end{equation}
These states are time independent (i.e. steady) states with respect to the reducing dynamics but they are, in general, not steady states with respect to both the original \textit{MESO} dynamics and the reduced dynamics.
Provided the thermodynamic potential $\Phi^{(Mm)}$ is specified, $x_{qeq}$ depends on $(\XX,\theta)$. In order $x_{qeq}$ could  play the role of the closure $\hat{x}(y)$,  $(\XX,\theta)$ have to be specified as functions of $y$. We shall call such specification a \textit{\textbf{constitutive relation}} $(\XX^{(CR)}(y),\theta^{(CR)}(y))$. The asymptotic solution  $x_{qeq}$ of (\ref{CRGENERIC}) with $\XX=\XX^{(CR)}(y)$ and $\theta =\theta^{(CR)}(y)$ will be denoted $x_{cl}(y)$. The lower index "cl" denotes "closure". We shall comment about constitutive relations at the end of this section. Now, we assume that the constitutive relations are known

The fundamental thermodynamic relation on \textit{meso level} implied by the fundamental thermodynamic relation (\ref{Mmcr}) on \textit{MESO level} is the following:
\begin{eqnarray}\label{Mmimpl}
\JJ^{(CR)}(\XX^{(CR)},\theta^{(CR)})&=& [\JJ(x)]_{x=x_{cl}} \\\label{thr}
y&=&[y(x)]_{x=x_{cl}}\nonumber \\
S^{(mM)*}(\XX^{(CR)},\theta^{(CR)})&=&[\Phi^{(Mm)}]_{x=x_{cl}}
\end{eqnarray}
The relation (\ref{Mmimpl})  is the specification of the reduced dynamics. The unspecified closure $\hat{x}(y)$ appearing in (\ref{Gdynnc}) is specified: $\hat{x}(y)=x_{cl}(\XX^{(CR)},\theta^{(CR)})$. The first line in (\ref{thr})
is the same as the first two lines in the fundamental thermodynamic relations (\ref{MIimpl}) and (\ref{Meimpl}) implied by $MICRO\rightarrow equilibrium$ thermodynamics. The second line in (\ref{thr}) is again the same as the third lines in (\ref{MIimpl}) and (\ref{Meimpl}). It is
a thermodynamic relations on \textit{meso} level implied by the reducing dynamics (\ref{CRGENERIC}). We emphasize that this relation is not implied by the reduced dynamics. As for the notation, we use the upper index $(mM)$ to denote  that the quantity is formulated on \textit{meso level} and is implied by dynamics on \textit{MESO level} (see the explanation of the notation in the text after (\ref{classftr})).
If we compare the  fundamental thermodynamic relation (\ref{Mmimpl}), (\ref{thr}) implied by $MESO\rightarrow meso$ with the fundamental thermodynamic relations (\ref{MIimpl}) and (\ref{Meimpl}) implied by $MICRO\rightarrow equilibrium$  and $MESO\rightarrow equilibrium$, we see that the new feature in  (\ref{Mmimpl}), (\ref{thr}) is the reduced dynamics  (\ref{Mmimpl}).
Indeed, the reduced dynamics in the approach to $MESO\rightarrow equilibrium$  is no dynamics and thus there is no need to specify it. If, on the other hand, we compare (\ref{Mmimpl}), (\ref{thr}) with standard investigations of reductions that put into focus only the reduced dynamics, the second line in (\ref{thr}) is new. It represents new thermodynamics implied by reducing dynamics. We also emphasize that the fundamental thermodynamic relation (\ref{Mmimpl}), (\ref{thr})  exists independently of whether the states in the reduced dynamics are steady or time dependent.

Finally, we return to the constitutive relations $(\XX^{(CR)}(y),\theta^{(CR)}(y))$ introduced in the text after Eq.(\ref{qeq}).  First, we note that in the context of $MESO\rightarrow equilibrium $ investigations in Sections \ref{RDMee}, \ref{RDMMee} and \ref{RDMMee},  constitutive relations are specifications of $\omega^*$. This means that in constitutive relations arising in $MESO\rightarrow equilibrium $ investigations we are expressing the conditions under which the macroscopic systems under consideration are investigated. This is also true in the context of $MESO\rightarrow meso$ investigations but with two new features. First, the conditions involve now also external forces. The imposed external forces   are expressed in  some of the forces $X$. We shall denote them by $\XX^{(ext)}$.
Second, the remaining forces, denoted  $\XX^{(int)}$ must be specified, as well as the free energy $\Phi^{(0Mm)}$ by solving \textit{MESO} time evolution equation (\ref{Fdyn}) (i.e. constructing the phase portrait $\mathcal{P}^{MESO}$, and extracting from it slower changing  pattern representing the reduced \textit{meso} time evolution (see also discussion in Section \ref{RD}). This, of course, can be done only if (\ref{Fdyn}) is more specified. We have done it in the  example discussed in Section \ref{motex} and we shall make other illustrations in the next Section \ref{EX}. At this point we only mention that it is in the constitutive relations where the entropy $S^{(me)}$  enters the analysis. The entropy $S^{(Mm)}$ then typically becomes closely related to the production of the entropy $S^{(me)}$. Recall for example the Fourier and Navier-Stokes constitutive relations in fluid mechanics (see more in Section \ref{EX4}). They are expressed in terms of the conjugate state variables with respect to the local entropy that, in the classical fluid mechanics, plays the role of $S^{(me)}$. We have also seen the similar constitutive relations in
Section \ref{motex}.

We end this section with a few remarks. More comments and illustrations are then in  Section \ref{EX}.

The CR fundamental thermodynamic relation (\ref{Mmimpl}) is a relation involving only the state variables and the material parameters used on \textit{meso level} (\ref{Gdyn}). From the physical point of view, we expect that even if it is not directly related to Eq.(\ref{Gdyn}), it reflects  important properties of  solutions of (\ref{Gdyn}).  This is because both Eq.(\ref{Gdyn}) and the relation (\ref{Mmimpl}) address the same physics even if expressed on different levels of description. In particular,  we anticipate, on the physical ground, that the presence of bifurcations in solutions to (\ref{Gdyn}) expressing mathematically the presence of sudden changes in behavior (e.g. the onset of convection in the Rayleigh-B\'{e}nard system) is manifested in the CR fundamental thermodynamic relation  (\ref{Mmimpl}) as phase transitions. This anticipation is based on the experimentally observed  growth of fluctuations  in meso-measurements of macroscopic systems in situations in which their behavior changes dramatically
(we shall call them critical situations). From this observation we then conclude that in critical situations  the "distance" between \textit{meso} and \textit{MESO levels} diminishes and the critical behavior manifests itself on both \textit{meso}  and \textit{MESO levels}. Since the CR fundamental thermodynamic relation is inherited from the \textit{MESO level}, we expect to see the critical behavior also in it.

Even without specifying the CR thermodynamic potential $\Phi^{(Mm)}$, the fact that the constitutive relations arise from minimizing it implies Maxwell-type reciprocity relations (that, from the mathematical point of view, express   symmetry of the second derivatives of $\Phi^{(Mm)}$) among the thermodynamic fluxes and forces. In the case when the potential $\Phi^{(Mm)}$ is quadratic then these reciprocity relations become Onsager's relations. Examples of reciprocity relations that arise in chemical kinetics are worked out  in Section III B in \cite{GrmPavKlika}.

We ask now the following question. Given an externally driven macroscopic system, how do we find the CR thermodynamical potential $\Phi^{(Mm)}$ (see (\ref{PhiMm}))  corresponding to it? If we ask the same question, but with externally unforced macroscopic systems and with the thermodynamic potential $\Phi^{(Me)}$ (see (\ref{Phi1}) ) replacing  externally driven macroscopic systems and  the   CR thermodynamic potential $\Phi^{(Mm)}$, then the answer is the following. On the most macroscopic level (that is for externally unforced systems the level of classical equilibrium thermodynamics - see Section \ref{RDET}), the only way we can identify the thermodynamic potential $\Phi^{(ee)}$ is by making experimental observations (e.g. observation of the relation among $P,V,T$ and of the specific heat - see Section \ref{RDET}). The knowledge of $\Phi^{(ee)}$ on the level of the classical equilibrium thermodynamics can be then transferred, via the local equilibrium assumption, also to
the level of fluid mechanics.  On the level of kinetic theory,  we can take as the point of departure for the search of $\Phi^{(Me)}$ the Boltzmann kinetic equation (playing in this example the role of \textit{MESO} dynamics) and arrive (following Boltzmann) to the Boltzmann entropy by investigating properties of its solutions.
On the \textit{MICRO level}, it suffices to know all the mechanical interactions expressed in the energy $E^{MICRO}$ since the entropy on the \textit{MICRO level} is the universal Gibbs entropy (\ref{microftr}). On \textit{meso levels}, we may find $\Phi^{(Me)}$ by MaxEnt reduction from the \textit{MICRO level} (i.e. by maximizing the Gibbs entropy subjected to constraints expressing the mapping  from \textit{MICRO} to \textit{meso}  state spaces) and/or by relating  entropy to concepts arising in the information theory and the theory of probability. In Section \ref{EX2} we shall suggest a possible universal \textit{MICRO level} CR entropy.

\section{Reducing Dynamics: Examples}\label{EX}

Our objective in this section is to make a few comments and illustrations that will bring    a more concrete content to the investigation discussed in previous sections. As for the $MESO\rightarrow equilibrium$ passage, many very specific illustrations can be found in \cite{Grmadv} and references cited therein and in \cite{Obook}. In Section \ref{EX2}, we work out a new illustration in which \textit{MESO level} is replaced by \textit{MICRO level}. In Sections \ref{EX3} and \ref{EX4} we develop two simple examples illustrating the $MESO\rightarrow meso$ passage. In Section \ref{EX5} we use the \textit{CR meso-thermodynamics} to estimate volume fractions at which phase inversion occurs in a blend of two immiscible fluids. In Section \ref{EX1} we comment about reductions seen as a pattern recognition in phase portraits.

Before proceeding to  specific illustrations,   we shall comment about the physics and the experimental basis  of the general thermodynamics presented above. We begin with the classical equilibrium thermodynamics. This theory has emerged from an attempt to combine mechanics involved in large scale mechanical engines with heat. As it became clear later in the Gibbs equilibrium statistical mechanics, the heat is a manifestation, on the macroscopic scale,  of the mechanics on the microscopic (atomic) scale. To combine the  large scale mechanics with heat is to combine large scale mechanics with microscopic mechanics. The objective of the classical equilibrium thermodynamics is to incorporate the microscopic mechanics (or heat which, at the time when thermodynamics was emerging, was  a rather mysterious concept) into the large scale mechanics by ignoring all that is irrelevant to  our direct macroscopic interest.  In the classical equilibrium thermodynamics this has been  achieved by enlarging the concept of mechanical energy (by introducing a new type of energy, namely the internal energy) and by introducing the concept of entropy together with  the MaxEnt principle. The setting of the classical equilibrium thermodynamics is thus a  two-level setting:  one level  (macroscopic) is  of our direct interest and the other  (microscopic)  is not of  our direct interest. We cannot however  completely ignore it since  it influences what happens on the macroscopic level. It is the concept of entropy that on the macroscopic level represents all from the microscopic level that is important for describing the behavior that directly interests us. All the other details involved on the microscopic level are ignored. The essence of the classical equilibrium thermodynamics is thus to provide a relation $MICRO\rightarrow macro$ between two levels of description.
Its experimental basis  consist of  observations showing that indeed the "minimalist" inclusion of the microscopic level offered by the classical equilibrium thermodynamics  leads to predictions that agree with the macroscopic experimental observations.

In the formulation of general thermodynamics we have extended the classical equilibrium thermodynamics by keeping  its two-level $MICRO\rightarrow macro$ setting but  we have replaced  the \textit{MICRO} and \textit{macro} levels with two general \textit{MESO} and \textit{meso} levels. Thermodynamics (including the classical equilibrium thermodynamics) is a
theory of theories or, in other words,  a metaphysics. The experimental basis of thermodynamics are meta-observations showing that behavior observed and well described on one level can also be observed  and well described on another level. Direct experimental observations, contrary to meat-observations, are observations made on a single level. They provide experimental basis of individual levels.
The concept of entropy can only be understood in the  two-level   $MESO\rightarrow meso$ viewpoint of thermodynamics. The often asked questions like for instance: does the entropy exist for driven systems, should be replaced with  the question: can the behavior of the driven system under investigation be described on two separate levels. If the answer to this latter question is yes then the answer to the former question is also yes. Since both well established levels are applicable, solutions of the time evolution on the level involving more details must approach solutions on the second level involving less details. The entropy is then the potential driving the approach.

In conclusion of the above comment about  the general thermodynamics  we note that
the very wide scope of thermodynamics (on the one hand it is a  metaphysics and on the other hand it is a very practically oriented engineering tool) is certainly one of the reasons for its attractiveness but it is also a reason    (at least one of the reasons) for unusually strong disagreements among its practitioners.

\subsection{Pattern recognition in the phase portrait, Chapman-Enskog method}\label{EX1}

An archetype example of \textit{MESO} time evolution equation (\ref{Fdyn}) is the Boltzmann kinetic equation. An archetype example of
$MESO\rightarrow meso$ investigation is the Chapman-Enskog analysis of the passage from the Boltzmann equation  to fluid mechanics. In this section we illustrate the pattern recognition viewpoint of reductions on the $MICRO\rightarrow MESO$ derivation of the Boltzmann equation and on the Chapman-Enskog method.

\subsubsection{MICRO $\rightarrow$ MESO introduction of the Boltzmann equation}\label{derBE}

We emphasize that our objective is not to derive rigorously the Boltzmann equation from \textit{MICRO} mechanics but only to illustrate how it can arise in the pattern recognition process in $\mathcal{P}^{MICRO}$.

In order to be able to recognize patterns in phase portraits $\mathcal{P}^{MICRO}$ we have to generate it  (or at least to obtain some pertinent information about it). Since $\mathcal{P}^{MICRO}$ is a collection of particle trajectories we have to find the trajectories, i.e. we have solve the \textit{MICRO} time evolution equations. It is important to realize that it is not the \textit{MICRO} vector field (i.e. the \textit{MICRO} time evolution equations) that is our starting point in in pattern recognition process but it the collection of trajectories that it generates (i.e. solutions to the \textit{MICRO} time evolution equations). In the case of a dilute  ideal gas (i.e. a macroscopic system composed of  particles that do not interact except for occasional binary collisions) the particle trajectories can be seen as a composition of straight lines (representing free particle motion) and two intersecting lines (representing binary collisions). Intersections of three (or more ) lines at one point (representing ternary (or higher order) collisions) are, due to the dilution, very rare and we therefore ignore them.
We choose  one particle distribution function $f(\rr,\vv)$ as the \textit{meso} state variable and try to recognize its time evolution (in particular the vector field generating it) that can be seen as a pattern in $\mathcal{P}^{MICRO}$.

We begin with the  straight line  $\rr\rightarrow \rr+\frac{\vv}{m}t$.  We can see this line  as a trajectory generated by $\dot{\rr}=\frac{\vv}{m}$ and $\dot{\vv}=0$. By $m$ we denote the mass of one particle and $t$ denotes the time. This particle time evolution induces  the time evolution $f(\rr,\vv)\rightarrow f(\rr-\frac{\vv}{m}t,\vv)$ in  one particle distribution functions. This time evolution is then  generated by the vector field
\begin{equation}\label{ffl1}
\frac{\partial f(\rr,\vv)}{\partial t}=-\frac{\partial}{\partial \rr}(\vv f(\rr,\vv))
\end{equation}
 We have thus arrived at the vector field representing the first feature of particle trajectories.

We turn now to the second feature, i.e. to two intersecting lines representing binary collisions.
Formally, we regarded this feature as  two straight lines, corresponding to momenta $(\vv_1,\vv_2)$,  meeting at the position with coordinate $\rr_1$ and continuing as two straight lines corresponding to momenta $(\vv'_1,\vv'_2)$.  The ingoing momenta $(\vv_1,\vv_2)$ and the outgoing momenta $(\vv'_1,\vv'_2)$ are related by the relations
\begin{eqnarray}\label{conBE1}
v_1^2+v_2^2&=&(v_1')^2+(v_2')^2\nonumber \\
\vv_1+\vv_2&=&\vv_1'+\vv_2'
\end{eqnarray}
expressing the mechanics of the collision. More details about the collision mechanics  (that would  make the relation between the ingoing and outgoing momenta one-to-one) are ignored and are not a part of   the second feature.  In terms of one particle distribution functions we express it therefore as the time evolution generated by the  gain-loss balance, or in other words, by considering $(\vv'_1,\vv'_2)\leftrightarrow(\vv_1,\vv_2)$ as a chemical reaction obeying  the constraint (\ref{conBE1}).
We shall see in Section \ref{EX2} that the vector field such gain-loss balance is given by
\begin{eqnarray}\label{intlin1}
\frac{\partial f(\rr_1,\vv_1)}{\partial t}&=&-\Xi^{(BE)}_{f^*(\rr_1,\vv_1)}\nonumber \\
&&=\int d2\int d1'\int d2'\widetilde{W}^{(BE)}(f(1')f(2')-f(1)f(2))
\end{eqnarray}
where $\Xi^{(BE)}$ is the dissipation potential, $\widetilde{W}^{(BE)}$ is a quantity appearing in it (see details in Section \ref{EX2} below), and $f^*(\rr_1,\vv_1)$ is a conjugate of $f(\rr_1,\vv_1)$ with respect to a entropy $S^{(BE)}(f)$  (i.e. $f^*=S^{(BE)}_f$).  With a  particular choice of these quantities,  the right hand side of (\ref{intlin1}) becomes the classical Boltzmann collision operator (see Section \ref{EX2}).

Both features of the particle phase portrait  $\mathcal{P}^{MICRO}$ are thus expressed in the time evolution of one particle distribution functions as the sum of the vector fields (\ref{ffl1}) and (\ref{intlin1}). The kinetic equation that we are obtaining in this way    is the Boltzmann kinetic equation.

The nonclassical formulation in which the Boltzmann equation is emerging  from our derivation
has several advantages. One of them is that the H-theorem (i.e. $\dot{S}^{(BE)}\geq 0$) is in it  manifestly visible. Another advantage  is that we have in fact derived  a generalization of the Boltzmann equation since we do not have to choose $S^{(BE)}(f)=-k_B\int d\rr\int d\vv f(\rr,\vv) \ln f(\rr,\vv)$.  The entropy  $S^{(BE)}(f)$ can be a more general potential $\int d\rr\int d\vv c(f(\rr,\vv))$, where $c$  is an unspecified but sufficiently regular concave function $\mathbb{R}\rightarrow\mathbb{R}$. With such more general entropy $S^{(BE)}(f)$ we  still have the H-theorem  $\dot{S}^{(BE)}\geq 0$ since, as we convince ourselves by a direct verification,  the time evolution generated by (\ref{ffl1}) does not change $S^{(BE)}(f)$. In addition, we also see that the step in the above introduction of the Boltzmann equation  where the time reversibility brakes and  the dissipation emerges is our ignorance of details of trajectories during binary collisions.

\subsubsection{Chapman-Enskog method}\label{CEmeth}

Let $\mathcal{P}^{MESO}$ and $\mathcal{P}^{meso}$ be the phase portraits corresponding to  the \textit{MESO} dynamics (\ref{Fdyn}) and the \textit{meso} dynamics (\ref{Gdyn}) respectively. Our problem is to recognize $\mathcal{P}^{meso}$ as a pattern inside of $\mathcal{P}^{MESO}$. While this viewpoint of the $MESO \rightarrow meso$ reduction does provide  a good intuitive understanding of the process, it does not provide a practical way to proceed. The archetype method offering such procedure is the Chapman-Enskog method (see e.g. \cite{GKHilb}).  This method  was originally developed for reducing the  Boltzmann kinetic equation to the  Navier-Stokes-Fourier hydrodynamic equations  but it can be applied to any $MESO \rightarrow meso$  passage. The pattern recognition process becomes in the context of the Chapman-Enskog method in the process of identifying a manifold $\mathcal{M}\subset M$ that satisfies the following two requirements: (i) $\mathcal{M}$  is in   one-to-one relation to $N$, (ii) $\mathcal{M}$ is quasi-invariant (i.e. $\mathcal{M}$ is "as much as possible" invariant with respect to the \textit{MESO} time evolution taking place on $M$).

We shall sketch below the geometrical essence of the method in three steps.
\\

\textit{Chapman-Enskog, Step 1}

By using an insight into the physics involved in the \textit{MESO} dynamics,  we write the \textit{MESO}  vector field $G$ as a sum of $G_0$, playing the dominant role, and $G_1$ that is seen as a perturbation (i.e. we write $G=G_0+G_1$).
The splitting of the vector field $G$  induces  then splitting of the search of the quasi invariant manifold $\mathcal{M}\subset M$ into two stages. A first approximation
$\mathcal{M}^{(0)}$ of $\mathcal{M}$ (called zero Chapman-Enskog approximation) is identified in the first stage (the second step in the Chapman-Enskog method) with neglecting $G_1$ (i.e. we consider $G=G_0$). In the second stage (the third step in the Chapman-Enskog method) the manifold   $\mathcal{M}^{(0)}$
is  deformed  into   $\mathcal{M}^{(1)}$   that is called a first Chapman-Enskog approximation of $\mathcal{M}$.

In the case of (\ref{Fdyn}) being the Boltzmann kinetic equation, $G_0$ is the Boltzmann collision term (since the pieces of the trajectories involving binary collisions are seen as being dominant in the phase portrait $\mathcal{P}^{MESO}$).
\\

\textit{Chapman-Enskog, Step 2}

In this step we identify $\mathcal{M}^{(0)}$. We define it as a manifold
on which the dominant vector field $G_0$ disappears (i.e. we solve the equation $[G_0]_{\mathcal{M}^{(0)}}=0$). The quantities that parametrize $\mathcal{M}^{(0)}$ are then chosen to be the \textit{meso} state variables $y$ expressed in terms of the \textit{MESO} state variables $x$. We thus obtain a mapping $\Pi:M\rightarrow N; x\mapsto y$. This mapping  subsequently induces a one-to-one  mapping $\Pi^{(\mathcal{M})}:\mathcal{M}^{(0)}\rightarrow N$. Next, the vector field $[G]_{\mathcal{M}^{(0)}}$ is  projected (by the projection induced by $\Pi^{(\mathcal{M})}$ ) on the tangent space of $\mathcal{M}^{(0)}$. We denote the projected vector field by the symbol  $G^{(0)}$. Finally, the vector field $G^{(0)}$  is projected (again by the projection induced by $\Pi^{(\mathcal{M})}$)  on the tangent space of  $N$. This is then   the vector field on $N$,  denoted by $g^{(0)}$ and called a zero Chapman-Enskog approximation of $G$ on $N$.

In the case of the \textit{MESO} dynamics (\ref{Fdyn}) being the Boltzmann kinetic theory, the mapping $\Pi$ is the standard mapping from  one particle distribution functions  to  hydrodynamic fields,  $\mathcal{M}^{(0)}$ is the manifold whose elements are local Maxwell  distribution functions, and $g^{(0)}$ is the right hand side of the  Euler (reversible and nondissipative) hydrodynamic equations.
\\

\textit{Chapman-Enskog, Step 3}

The first Chapman-Enskog approximation  $\mathcal{M}^{(1)}$ of $\mathcal{M}$ is found in this step.
We note that the manifold $\mathcal{M}^{(0)}$ is not an invariant manifold since the vectors  $[G]_{\mathcal{M}^{(0)}}$ do not
lie in the tangent spaces attached to $x_0\in \mathcal{M}^{(0)}$. We want to make it more invariant. We therefore deform
$\mathcal{M}^{(0)}$ into $\mathcal{M}^{(1)}$ $( x_0\mapsto x_1)$  in such a way  that $G^{(1)}\equiv [G]_{\mathcal{M}^{(0)}}$,
where $G^{(1)}$ is the vector field $G$ attached to the points $x_1$ and projected on $\mathcal{M}^{(1)}$. We note that the
manifold $\mathcal{M}^{(1)}$ is still not invariant (since, in general,  $[G]_{\mathcal{M}^{(1)}}\neq [G]_{\mathcal{M}^{(0)}}$)
but it is expected to  be "more" invariant than $\mathcal{M}^{(0)}$ since the vector field $G_1$ is just a perturbation of $G_0$
(and consequently the deformation $\mathcal{M}^{(0)}\rightarrow \mathcal{M}^{(1)}$ is small). The vector field $g^{(1)}$ projected on $N$
is the first Chapman-Enskog approximation of $G$.

In the case of (\ref{Fdyn}) being the Boltzmann equation,  the vector field   $g^{(1)}$ is  the right hand side of the Navier-Stokes-Fourier
(irreversible and dissipative) hydrodynamic equations.

If both  \textit{MESO} dynamics (\ref{Fdyn}) and  \textit{meso} dynamics (\ref{Gdyn}) are known and well established (i.e. they both have emerged
from direct derivations involving \textit{MESO} measurements and \textit{meso} measurements respectively) then  the Chapman-Enskog type derivation
of (\ref{Gdyn}) from (\ref{Fdyn}) brings an additional information.  First, the domain of applicability of (\ref{Gdyn}) inside of the domain
of applicability of (\ref{Fdyn}) is identified, and second,  mapping $\xi \mapsto \zeta$ emerges (i.e. the material parameters $\zeta$ with
 which individual features of the systems under consideration are expressed  on the \textit{meso level} become functions of the material parameters $\xi$
 used for the same purpose on the \textit{MESO level}).

Examples of applications of the Chapman-Enskog method in many types of mesoscopic dynamics (including for instance the dynamics describing
chemical reactions) can be found in \cite{Gorbbook}, \cite{GKHilb}.

We make now three additional comments about the Chapman-Enskog method.

\textit{Comment 1}

We note that there is no thermodynamics in the Chapman-Ensog method. How can we bring it to it? Following the viewpoint of thermodynamics presented in previous sections, we have to turn attention not only to the pattern (i.e. in this case to the submanifold  $\mathcal{M}_1$)   but also to the reducing time evolution bringing $x\in M$ to it. As we have seen, the reducing tome evolution is generated by a potential so, at leat, we should try to identify the potential. As for the manifold  $\mathcal{M}_0$, the reducing time evolution is the Boltzmann equation without $G_1$ (i.e. Eq.(\ref{intlin1}) and the potential is obviously  the  Boltzmann entropy. Indeed,  $\mathcal{M}_0$ can be obtained by MaxEnt reduction of the Boltzmann entropy. The manifold on which $S^{(BE)}(f)$ reaches its maximum subjected to constraints representing the fluid mechanics fields expressed in terms of the one particle distribution function  is exactly the submanifold  $\mathcal{M}_0$.
For example in \cite{Gorbbook}, this is  the way  the submanifold  $\mathcal{M}_0$ is introduced.  The second step in the Chapman-Enskog method can be thus seen as  a part of  the investigation of reducing dynamics.

Can we follow this path and interpret thermodynamically also the third step in the Chapman-Enskog method (i.e. the deformation of $\mathcal{M}_0$ to $\mathcal{M}_1$)? In order to make such interpretation, we  look for a potential $S^{(BE)}_1(f)$, that satisfies the following properties: (i) $S^{(BE)}_1(f)$ is a deformation of $S^{(BE)}(f)$, (ii) its maximum is reached at $\mathcal{M}_1$, and (iii) it generates the time evolution in which  $\mathcal{M}_1$ is approached (similarly as $S^{(BE)}(f)$ generates the time evolution  $\mathcal{M}_0$ is approached. The partial results related to  this problem that are reported in Section 4.2 of \cite{Grmadv} indicate that the potential obtained in this way is indeed the CR-entropy generating the reducing time evolution that is  involved in the passage from kinetic theory to fluid mechanics.

\textit{Comment 2}

For the kinetic equation (\ref{intlin1}) the submanifold $\mathcal{M}_0$ is an invariant manifold. For the full Boltzmann kinetic equation (i.e. kinetic equation combining (\ref{ffl1}) and (\ref{intlin1})) neither $\mathcal{M}_0$ nor $\mathcal{M}_1$ are invariant manifolds. In fact, as it has been shown by Grad in \cite{Grad}, and Desvillettes and Villani in \cite{Vill}, the only invariant manifold is the manifold $\mathcal{M}_{eq}$ of equilibrium sates (i.e. time independent solutions of the full Boltzmann equation). Both manifolds  $\mathcal{M}_0$ and  $\mathcal{M}_1$ are quasi-invariant manifolds. Grad,  Desvillettes and Villani have proven that solutions to the full Boltzmann equation may come very close to $\mathcal{M}_0$ and  $\mathcal{M}_1$ (that is why we can call  these manifolds  quasi-invariant manifolds) but they never fall on neither of them. They only fall eventually on the submanifold $\mathcal{M}_{eq}$  that is a submanifold of both $\mathcal{M}_0$ and  $\mathcal{M}_1$.

\textit{Comment 3}

Reduction to the submanifold $\mathcal{M}_0$ results in the Euler fluid mechanics equations and reduction to its deformation $\mathcal{M}_1$ results in the Navier-Stokes-Fourier fluid mechanics equations. Both these fluid mechanics  equations are particular realizations of (\ref{GENERIC}) and thus both are physically meaningful.

In principle,  it is possible to continue the deformations of $\mathcal{M}_0$. Similarly as
we have made the deformation $\mathcal{M}_0\rightarrow\mathcal{M}_1$ we can make the next deformation $\mathcal{M}_1\rightarrow \mathcal{M}_2$. In other words, we can proceed to  the second Chapman-Enskog approximation. Will be the resulting reduced time evolution again physically meaningful (in the sense that it will be  a particular realization of (\ref{GENERIC}))? Experience collected in the investigations of higher order Chapman-Enskog approximations (e.g. the investigation of the linearized Boltzmann equation in \cite{GRHELV}) seems to indicate that the answer to this question is negative.

\subsection{MICRO $\rightarrow$  equilibrium reducing dynamics}\label{EX2}

Many illustrations of GENERIC equation (\ref{GENERIC}) in kinetic theory,  fluid mechanics, and solid mechanics of simple and complex fluids  can be found in \cite{Obook}, \cite{Grmadv} and references cited therein. In this section we develop an additional new illustration. We return to the Gibbs equilibrium statistical mechanics presented in Section \ref{RDMee} and ask the following question. What is the GENERIC time evolution of the N-particle distribution function $f(\rr,\vv)$ (i.e. the time evolution governed by  (\ref{GENERIC}) with $x=f_N(1,...,N)$)  that makes the maximization of the Gibbs entropy (see (\ref{microftr})) postulated MaxEnt  in the point (iv) of the \textit{MICRO $\rightarrow$ equilibrium  Postulate II}?  We first introduce one such time equation and then discuss its possible  non uniqueness. We use hereafter a shorthand notation $1\equiv (\rr_1,\vv_1), 2\equiv (\rr_2,\vv_2),...)$.

We look for a dissipation potential $\Xi^{(N)}$  which brings N-particle distribution functions $f_N(1,...,N)$ to the  Gibbs distribution $(f_N)_{eq}$ (i.e. to $f_N$ for which the thermodynamic potential $\Phi$ given in (\ref{Phi}) reaches its minimum). In other words, we look for $\Xi^{(N)}$  for which solutions to
\begin{equation}\label{Nkin}
\frac{\partial f_N(1,2,...,N)}{\partial t}=-\Xi^{(N)}_{f^*_N(1,2,...,N)}
\end{equation}
approach, as $t\rightarrow\infty$,   $(f_N)_{eq}$. Inspired by the dissipation potential arising in the Guldberg-Waage chemical kinetics (see \cite{Grmchem}) and the dissipation potential generating the Boltzmann collision integral  \cite{GrB}, we propose
\begin{eqnarray}\label{XiN}
\Xi^{(N)}&=&\int d1...\int dN\int d1'...\int dN'W^{(N)}(f_N,1,...,N,1',...,N')\nonumber \\
&&\times\left(e^{\frac{1}{2}X^{(N)}}+e^{-\frac{1}{2}X^{(N)}}-2\right)
\end{eqnarray}
where the thermodynamic forces are given by
\begin{equation}\label{Nforce}
X^{(N)}=\frac{1}{k_B}(f^*_N(1,2,...,N)-f^*_N(1',2',...,N')),
\end{equation}
$f^*_N=S_{f_N}$, $S(f_N)$ is the Gibbs entropy (\ref{microftr}),
\begin{equation}\label{12}
(1,2,...,N)\rightleftarrows (1',2',...,N')
\end{equation}
are one-to-one transformations in which the microscopic energy $E^{MICRO}(1,...N)$ remains constant, i.e.
\begin{equation}\label{12con}
E^{MICRO}(1,2,...,N)=E^{MICRO}(1',2',...,N'),
\end{equation}
and $W^{(N)}\geq 0$ are nonegative material parameters  that are  different from zero ($W^{(N)}\neq 0$)  only if the constraint (\ref{12con}) holds and $W$ is symmetric with respect to $(1,2,...,N)\rightarrow (1',2',...,N')$.

In the Guldberg-Waage chemical kinetics (see \cite{Grmchem}), the transformation (\ref{12}) is interpreted as a chemical reaction. We shall demonstrate below that the Boltzmann collision operator is the right hand side of (\ref{Nkin}) with $N=$, dissipative forces $X^{(1)}$ given in (\ref{X1}) and the transformation (\ref{12}) given in (\ref{binreacx}) and (\ref{conBE}).

Before proving that solutions to (\ref{Nkin}) approach $(f_N)_{eq}$, we write the time evolution equation (\ref{Nkin}) explicitly. With the Gibbs entropy (\ref{microftr}), Eq.(\ref{Nkin}) takes the form
\begin{eqnarray}\label{Nkinexp}
\frac{\partial f_N(1,2,...,N)}{\partial t}&=&-\Xi^{(N)}_{f^*_N(1,2,...,N)}\nonumber \\
&&=\int d1'...\int dN'\widetilde{W}^{(N)}
(f_N(1',...,N')-f_N(1,...,N))\nonumber \\
\end{eqnarray}
where $\widetilde{W}^{(N)}=\frac{W^{(N)}}{2k_B(f_N(1,...,N)f_N(1',...,N'))^{\frac{1}{2}}}$.

The Legendre transformation  $\Theta^{(N)}(J)$ of $\Xi^{(N)}(X)$ is
\begin{eqnarray}\label{Theta}
\Theta^{(N)}(J)&=&2\int d1...\int dN\int d1'...\int dN' W\nonumber \\
&&\times\left[\hat{J}\ln\left(\hat{J}+\sqrt{1+(\hat{J})^2}\right)-\left(\sqrt{1+(\hat{J})^2} -1\right)\right]
\end{eqnarray}
where $\hat{J}=\frac{J}{W}$.

Now we prove that solutions to (\ref{Nkin}) (or (\ref{Nkinexp})) approach, as $t\rightarrow \infty$, the  Gibbs distribution $(f_N)_{eq}$.
First, we see that the right hand side of (\ref{Nkin}) equals zero if $X=0$. In view of  (\ref{12con})), equation $X=0$ is solved by $f_N=(f_N)_{eq}$. Since the thermodynamic potential $\Phi$ plays the role of the Lyapunov function for the approach to $(f_N)_{eq}$ (see (\ref{asGEN})), we  see that solutions to (\ref{Nkin}) (which takes the form (\ref{Nkinexp}) provided  the entropy is the Gibbs entropy,) approach, as $t\rightarrow \infty$, the  Gibbs distribution $(f_N)_{eq}$. The time evolution governed by (\ref{Nkin}) indeed brings macroscopic systems to states investigated in the Gibbs equilibrium statistical mechanics (see Section \ref{RDMee}. The dissipation potential $\Xi^{(N)}$ given in (\ref{XiN}) (or equivalently its Legendre transformation $\Theta^{(N)}$ given in (\ref{Theta}))  can be therefore regarded as  the universal CR-entropy on \textit{MICRO level}  similarly as the Gibbs entropy (\ref{microftr}) is the universal entropy on \textit{MICRO level}. Consequently, we can  find CR-entropies on \textit{meso levels}  similarly  as we can find  entropies on \textit{meso levels}. We can either try to extract them from the time evolution
(generating  $meso \rightarrow equilibrium$  passage - in the case of entropy, or $MESO \rightarrow  meso$ passage - in the case of CR-entropy) or we can attempt to reduce them (by MaxEnt) from the universally valid expressions (for the Gibbs entropy (\ref{microftr}) in the case of entropy and for the CR-entropy (\ref{XiN}) in the case of CR-entropy).

We  now show that  the dissipation potential (\ref{XiN}) can be seen as a natural extension of the dissipation potential generating the Boltzmann collision operator arising  in one particle kinetic theory. At the end of this section  we then investigate other possible vector fields that can describe approach to equilibrium.

The time evolution equation (\ref{Nkin}) has a well defined meaning  for any $N\geq 2$. For $N=1$, i.e. for the level of one particle kinetic theory, we cannot make  the transformation (\ref{12}) and we cannot therefore directly use (\ref{Nkin}). In order to be able to introduce  transformations of the type (\ref{12}) in one particle kinetic theory,  we  need a partner. We shall denote the  coordinates of the particle by   $(\rr_1,\vv_1)$ and  of its partner by  $(\rr_2,\vv_2)$. From the physical point of view, we  regard   the transformation (\ref{12}) in the context of one particle kinetic theory as a binary collision between the particle and its partner. We therefore write (\ref{12})  in the form
\begin{equation}\label{binreacx}
(\vv_1,\vv_2)\rightleftarrows (\vv'_1,\vv'_2),
\end{equation}
with the constraint
\begin{eqnarray}\label{conBE}
v_1^2+v_2^2&=&(v_1')^2+(v_2')^2\nonumber \\
\vv_1+\vv_2&=&\vv_1'+\vv_2'
\end{eqnarray}
replacing  the constraint (\ref{12con}).
The binary collision  are assumed to take place at a fixed point with the spatial coordinate $\rr_1$. The  constraints (\ref{conBE}) express the conservation of energy and   momentum in the collisions.

With the new transformation (\ref{binreacx})  we then  replace the thermodynamic force (\ref{Nforce}) by
\begin{equation}\label{X1}
X^{(1)}=\frac{1}{k_B}(f^*(\rr_1,\vv_1)+f^*(\rr_1,\vv_2)-f^*(\rr_1,\vv'_1)-f^*(\rr_1,\vv'_2)),
\end{equation}
The time evolution equation (\ref{Nkin}) takes now the form
\begin{eqnarray}\label{intlin}
\frac{\partial f(\rr_1,\vv_1)}{\partial t}&=&-\Xi^{(1)}_{f^*(\rr_1,\vv_1)}\nonumber \\
&&=\int d2\int d1'\int d2'\widetilde{W}^{(1)}(f(1')f(2')-f(1)f(2))
\end{eqnarray}
where $\widetilde{W}^{(1)}=\frac{W^{(1)}}{2k_B(f(1)f(2)f(1')f(2'))^{\frac{1}{2}}}$, $W^{(1)}$ is symmetric with respect to the transformation  $\vv_1 \leftrightarrows \vv_2, \, \vv'_1 \leftrightarrows \vv'_2$ and $(\vv_1,\vv_2)\rightleftarrows (\vv'_1,\vv'_2)$. Equation (\ref{intlin})
is the Boltzmann kinetic equation without the free flow term (i.e. the right hand side of (\ref{intlin}) is the Boltzmann collision operator - see more in \cite {GrPhysD}, \cite{Grmadv}).

Since the Boltzmann collision dissipation appears to be  essentially a special case of the dissipation introduced  in (\ref{Nkin}),  we can indeed regard the dissipation potential $\Xi^{(N)}$ in (\ref{XiN}) as a natural extension of the dissipation potential generating the classical Boltzmann  binary collision dissipation.

There is however an interesting difference between the Boltzmann dissipation in (\ref{intlin}) and the dissipation appearing in (\ref{Nkin}). The former is weaker than the latter since the Boltzmann dissipation drives solutions to local equilibrium while the dissipation appearing in (\ref{Nkin}) drives solutions to the total equilibrium. In general,
we say that the dissipation generated by  the vector field $\left(vector\,field\right)_1$ is stronger than the dissipation generated by the vector field $\left(vector\,field\right)_2$ if the inequality $\dot{\Phi}\leq 0$ holds for both vector fields but $\mathcal{M}_1\subset \mathcal{M}_2$. By   $\mathcal{M}_i$  we denote  the manifold whose elements are  states  approached as
$t\rightarrow\infty$ in the time evolution generated by the vector field $\left(vector\,field\right)_i$; $i=1,2$.

Let us now consider two vector fields: one is given by the right hand side of (\ref{intlin}) and the other by the right hand side of
\begin{equation}\label{ffl}
\frac{\partial f(\rr,\vv)}{\partial t}=-\frac{\partial}{\partial \rr}(\vv f(\rr,\vv))
\end{equation}
From the physical point of view, (\ref{ffl})  is  one particle kinetic equation representing a gas of completely noninteracting particles with no collisions. It is a (continuity) Liouville equation corresponding to the particle dynamics $\dot{\rr}=\vv; \,\dot{\vv}=0$.
We note that the vector field (\ref{ffl}) is nondissipative (i.e. Eq.(\ref{ffl}) implies  $\dot{\Phi}=0$) while, as we have shown above, the vector field (\ref{intlin}) is dissipative (i.e. Eq.(\ref{intlin}) implies $\dot{\Phi}\leq 0$) and the manifold $\mathcal{M}_{leq}$ corresponding to it is the manifold composed of local Maxwell distribution functions.
Grad in \cite{Grad}, and Desvillettes and Villani (in full generality)  in \cite{Vill},   have  proven  the following result. The manifold $\mathcal{M}_{teq}$ corresponding to the sum of the vector fields (\ref{ffl}) and (\ref{intlin}) (i.e. to the vector field appearing in the Boltzmann kinetic equation) is the manifold composed of total Maxwell distribution functions. Since  $\mathcal{M}_{teq}\subset  \mathcal{M}_{leq}$,
we see that we can make dissipation generated by a vector field stronger just by adding to it an appropriate  nondissipative vector field.

This result, if transposed to the setting of N-particle dynamics for  $N\geq 2$, indicates that vector fields with weaker dissipation than the vector field (\ref{Nkinexp}) can  possibly still drive solutions to the Gibbs equilibrium distribution function $(f_N)_{eq}$ provided the nondissipative vector field arising in the Liouville N-particle equation is added to them. What could be the vector fields that have a weaker dissipation than the vector field (\ref{Nkinexp})? One way to construct them  is to keep the dissipation potential (\ref{XiN}), to keep the  thermodynamic force (\ref{Nforce}), to keep the interaction (\ref{12}) but to introduce  stronger constraints so that solutions to $X^{(N)}=0$ form a smaller manifold. For example, we can replace the constraint (\ref{12con}) with the constraint: $(1',...,N')$ is just a reordering of $(1,...,N)$. In the case of $N=2$, this constraint becomes $(1',2')=(2,1)$. For such dissipation vector field the manifold $\mathcal{M}$ of states approached as $t\rightarrow\infty$
is the manifold of symmetric distribution functions. The time evolution generated by this vector field drives distribution functions to symmetric distribution functions. The time evolution is making the symmetrization.  The following question then arises. Is this symmetrization dissipation, if combined with the nondissipative Liouville vector field,  strong enough to drive solutions to the Gibbs equilibrium distribution $(f_N)_{eq}$? If the answer is negative (as it is probably the case), the next question is then to identify the dissipation potential with the weakest possible dissipation that, if combined with the Liouville vector field, does drive solutions to $(f_N)_{eq}$. In this paper we leave these questions unanswered.

Before leaving this section we note that another interesting variation of the constraint (\ref{conBE}) arises in the investigation of
granular gases (i.e. gases composed of particles of  macroscopic size). In this case the collisions are inelastic so that the first
line in (\ref{conBE}) is missing. Granular gas is an interesting example of an externally driven macroscopic system that can naturally be
investigated on the level of kinetic theory (see e.g. \cite{granular}). Its thermodynamic investigation could be then based on the CR-thermodynamic
potential with the dissipation potential $\Xi^{(1)}$ given in (\ref{XiN}), and with two imposed forces:
thermodynamic force  $X^{(1)}$ (see (\ref{X1})) representing the inelastic  collisions, and a
mechanical force $X^{(mech)}$ representing for example shaking.

\subsection{equilibrium $\rightarrow$ equilibrium (imposed temperature)}\label{EX3}

The external force in this example is the energy exchange with  thermal bath that is kept at a constant temperature $\mathfrak{T}$. In this case the externally driven macroscopic system evolves to an equilibrium state. This means that in this investigation of the  $MESO \rightarrow meso$  passage the \textit{meso level} is the  equilibrium level. The difference between the $MESO \rightarrow equilibrium$ passage investigated in Section \ref{RDMMde} and  the  $MESO \rightarrow equilibrium$ passage investigated in this example is that the state variables $y_{eq}$ at the equilibrium level are not $(E,V,N)$  (as in Section \ref{ET}) but $(E^*,V,N)$, where $E^*=S_E=\frac{1}{T}$ is the conjugate of $E$.

The CR thermodynamic potential $\Psi$ driving the  evolution is, in this example, the thermodynamic potential (\ref{Phi1}) with $T=\mathfrak{T}$, i.e. $\Phi(x,\mathfrak{T},\mu)=-S(E,V,N)+\frac{1}{\mathfrak{T}}E-\frac{\mu}{\mathfrak{T}}N$. The CR-GENERIC equation (\ref{CRGENERIC}) is, in this example,  the GENERIC time evolution (\ref{GENERIC}) with $T=\mathfrak{T}$ and with the degeneracies of $L$ and $\Xi$ that guarantee the mass conservation (\ref{consNGEN}) but not the energy conservation (\ref{consEGEN}).  The resulting fundamental thermodynamic relation is the Legendre transformation $(E,N,V)\rightarrow(\frac{1}{T},N,V)$ of the fundamental thermodynamic relation $S=S(E,V,N)$ implied by the GENERIC reducing time evolution discussed in Section \ref{RDMMde}.

The above analysis becomes particularly interesting if we choose the \textit{MESO level} to be the equilibrium level with state variables $(E,V,N)$. In this case of $MESO \rightarrow meso$ passage both \textit{MESO}  and \textit{meso levels} are \textit{equilibrium levels}. They differ only in state variables: on \textit{MESO level} the state variables are $(E,N,V)$ and on \textit{meso level} $(E^*,V,N)$
The reducing time evolution in this $equilibrium \rightarrow equilibrium$ passage is the  time evolution making the Legendre transformation  $(E,N,V)\rightarrow(\frac{1}{T},N,V)$. The thermodynamic potential generating it is
\begin{equation}\label{phie}
\Phi(E,N,V,\mathfrak{T})=-S(E,N,V)+\frac{1}{\mathfrak{T}}E
\end{equation}
and the time evolution equation (\ref{CRGENERIC}) becomes
\begin{equation}\label{ee}
\dot{E}=-[\Xi_X(E,X)]_{X=-S_E+\frac{1}{\mathfrak{T}}}
\end{equation}
that, if we choose the quadratic dissipation potential  $\Xi(E,X)=\frac{1}{2}\Lambda X^2$, where $\Lambda>0$ is a material parameter,  becomes $\dot{E}= -\Lambda (-S_E+\frac{1}{\mathfrak{T}})$.

\subsection{Cattaneo $\rightarrow$ Fourier (imposed   temperature gradient)}\label{EX4}

In this example
the external force is  an imposed temperature gradient. We  denote it by the symbol $\nabla\frac{1}{\mathfrak{T}}$. This force prevents  approach to \textit{equilibrium level}. The most macroscopic level (i.e. the level with least details) on which macroscopic systems subjected to  temperature gradient can be described is the level of fluid mechanics (we shall call it hereafter FM-level) on which the state variables are: $x=(\rho(\rr),\uu(\rr),e(\rr))$, where $\rr$ is the position vector, $\rho(\rr)$ is the mass field (mass per unit volume at $\rr$), $\uu(\rr)$ is the momentum field, and $e(\rr)$ the energy field. In this example we shall limit ourselves only to the state variable $e(\rr)$. All other state variables are assumed to be already at equilibrium. In this setting we now  investigate the passage $MESO \rightarrow FM$.

First, we recall that in the  absence of the  imposed temperature gradient (i.e. if $\nabla\frac{1}{\mathfrak{T}}=0$, the macroscopic systems under consideration will approach to \textit{equilibrium level} and we can therefore consider the level of fluid mechanics (called FM-level) as \textit{MESO level} and investigate  the passage $FM \rightarrow equilibrium$. The GENERIC equation (\ref{GENERIC}) representing this passage is well known and can be found for example in \cite{Grmadv}.

Now we switch on the external force (i.e. $\nabla\frac{1}{\mathfrak{T}}\neq 0$) and investigate the $MESO \rightarrow FM$  passage.
We proceed to find the CR-GENERIC equation representing it. The state variables that evolve in the reducing time evolution is the  heat flux $J^{(h)}$. The CR relation is the Fourier  constitutive relations: $X^{(h)}_i=\nabla_i\frac{1}{S^{(leq)}_E}$,  where $S^{(leq)}(\rho(\rr),\uu(\rr),e(\rr))$ is the local equilibrium entropy on the \textit{FM level}.
We use hereafter the indices $i=1,2,3;\, j=1,2,3$ and the summation convention.

The CR thermodynamic potential in this example is
\begin{equation}\label{CRFM}
\Phi^{(MFM)}(J^{(h)}; (\nabla\frac{1}{\mathfrak{T}}))=-S^{(0MFM)}(J^{(h)})+(\nabla\frac{1}{\mathfrak{T}})_iJ^{(h)}_i
\end{equation}
In the CR-GENERIC equation (\ref{CRGENERIC}) we  neglect the Hamiltonian time evolution (i.e. we put $\mathcal{L}\equiv 0$) and, for the sake of simplicity, choose  the dissipation potential
$\Xi^{(FMe)}(J^{(h)}),X^{(h)})=\frac{1}{2}\int d\rr (\Lambda^{(h)}X_i^{(h)}X_i^{(h)}$, where $\Lambda^{(h)}>0$ is a material parameter.  With these specifications the CR-GENERIC equation (\ref{CRGENERIC}) becomes
\begin{equation}\label{delt}
\dot{J}^{(h)}_i=-\Lambda^{(h)}(-S^{(0FMe)}_{J^{(h)}_i}+(\nabla\frac{1}{\mathfrak{T}})_i)
\end{equation}
The fundamental thermodynamic relation on the \textit{FM level}  implied by the \textit{MESO level} CR-GENERIC equation (\ref{delt}) is
\begin{equation}\label{frFM}
S^{(FMM)*}(e(\rr);(\nabla\frac{1}{\mathfrak{T}}))=\left[\Phi^{(MFM)}(J^{(h)}; (\nabla\frac{1}{\mathfrak{T}}))\right]_{S^{(0MFM)}_{J^{(h)}_i}=(\nabla\frac{1}{\mathfrak{T}})_i}
\end{equation}
where $\Phi^{(MFM)}$ is the CR thermodynamic potential (\ref{CRFM}).

We note that Eq.(\ref{delt}) is the well known Cattaneo equation \cite{Cattaneo} provided the imposed external force $\nabla\frac{1}{\mathfrak{T}}$ is replaced by $\nabla\frac{1}{T}$, where $T(\rr)$ is the local temperature. There is however an important difference between the role it plays in extended thermodynamic theories in \cite{MullRugg}, \cite{Joubook} and in this paper. In the context of our investigation it is the equation describing approach of the Cattaneo  extended fluid dynamics (playing the role of \textit{MESO level}) to the classical fluid mechanics with the Fourier constitutive relation (playing the role of \textit{meso level}). The Cattaneo time evolution  driven by the CR thermodynamic potential (\ref{CRFM}), implying the  CR fundamental thermodynamic relation (\ref{frFM}) on the level of classical fluid mechanics, describes $MESO\rightarrow meso$ passage. In the extended theories investigated in  \cite{MullRugg}, \cite{Joubook}   the Cattaneo equation is the equation arising in the $MESO\rightarrow equilibrum$ passage. It is
just an extra equation (governing the time evolution of the extra state variable and coupled to the other time evolution equations) in the set of extended fluid mechanics equations whose solutions are required to approach equilibrium states. This means that the physical systems under investigation in  \cite{MullRugg}, \cite{Joubook} are externally unforced.

\subsection{Thermodynamics of immiscible blends; phase inversion}\label{EX5}

In this section we  apply  CR-thermodynamics to immiscible blends. We recall that   the  extension of the classical equilibrium thermodynamics of  single component macroscopic systems to multicomponent miscible blends  led Gibbs  to the   completion of the mathematical  formulation of  equilibrium thermodynamics. Further extensions to immiscible blends require to leave the realm of  equilibrium thermodynamics and to enter CR-thermodynamics. Imposed external forces (e.g. imposed flows in the mixing process) prevent approach to
equilibrium states. Moreover, extra  variables addressing  morphology of the interfaces among the components are needed to characterize their states. We shall not attempt in this paper to make a systematic investigation of CR-thermodynamics of immiscible blends. We shall concentrate only on one particular problem and use thermodynamics to investigate  it.

The immiscible blend that we consider is composed of two immiscible fluids (component "1", and component "2").
The problem that we investigate is phase inversion.  Let initially the component "1" form  a continuous phase in which the component "2" is dispersed. This means that the component "2" resides inside drops encircled completely by the component "1". Every two points in the component "1" can be joined by a line that lies completely inside the component "1". We shall now increase the amount of the component "2". We anticipate that at some volume fraction $\phi_2$ of the second component the roles of the two components change, the second component becomes the  continuous phase and the first component becomes the dispersed phase. At the critical state at which the change occurs both components form a continuous phase. The morphology at the critical state is called a co-continuous  morphology. The problem that we want to investigate is to estimate the critical value of $\phi_2$ as a function of the properties of the components (as for instance the viscosity, elasticity etc.) and of the blending conditions (i.e. the externally imposed forces). The co-continuous morphology is in particular very important in applications involving blends of polymer melts (see e.g. \cite{Favis}).

In order to have a specific example in mind, we can think of  immiscible blends of oil and water. If water is the continuous phase then the blend is milk, if oil is the continuous phase then the blend is butter.

One way to approach the problem of phase inversion is by attempting to formulate a dynamical model of immiscible blends.
In this paper we  shall not take this route. We  turn directly to thermodynamics.   The concept with which we begin is the CR thermodynamic potential (\ref{PhiMm}). Next, we  regard phase inversion as phase transition. In view of the  comment that we made at the end of Section \ref{GF}, this means that at the point of phase inversion
\begin{equation}\label{Phinver}
\Phi^{(mM)}_{1/2}=\Phi^{(mM)}_{2/1}
\end{equation}
where
$\Phi^{(mM)}_{i/k}$ is the CR thermodynamic potential when the component "i" forms the continuous phase and the component "k" the disperse phase. This is  the equation that  answers our question.
It remains now only to specify the CR thermodynamic potential $\Phi^{(mM)}$.

We shall limit ourselves  to  most simple specifications that nevertheless illustrate the power of CR-thermodynamics. Following (\ref{Psi}), we write
\begin{equation}\label{Pimbl}
\Phi^{(mM)}= -S^{(0mM)}+\frac{1}{T_0}W
\end{equation}
where we consider $S^{(0mM)}$ to be simply  the local entropy production due to the presence of the dispersed phase,   $W$  the work involved in elastic deformations of the dispersed droplets, and $T_0$  the temperature of the blend.

Our problem now is to estimate $S^{(0Mm)}$ and $W$ in the mixture in which $\phi_1$ and $\phi_2$ are not too different.
Let  the component "1", forming  the continuous phase, be a fluid of viscosity $\eta_1$.  The main contribution to the entropy production $S^{(0Mm)}_{1/2}$  comes from the flow in the continuous phase "1". $S^{(0mM)}_{1/2}$  is thus
$\alpha_2\eta_1\phi_2 \dot{\gamma}^2$, where $\alpha_2$ is a parameter depending on the shape of the inclusion  and $\dot{\gamma}$  is the absolute value of the shear rate. The main contribution to the work $W_{1/2}$ is assumed to come also from  the continuous phase, i.e. $W_{1/2} \sim  H_1D_1^2\dot{\gamma}$, where $H_1$ is the elastic constant and  $D_1$ is the deformation displacement of the matrix). We therefore obtain
\begin{equation}\label{P1}
S^{(0mM)}_{1/2}=\alpha_2\phi_2\eta_1\dot{\gamma}^2;\,\,\,
S^{(0mM)}_{2/1}=\alpha_1\phi_1\eta_2\dot{\gamma}^2
\end{equation}
and
\begin{equation}\label{W}
W_{1/2}=\phi_1H_1D_1^2\dot{\gamma};\,\,\,
W_{2/1}=\phi_2H_2D_2^2\dot{\gamma}
\end{equation}
By inserting these relations to (\ref{Pimbl}) and (\ref{Phinver}) we arrive finally at the estimate of the critical volume fraction
\begin{equation}\label{result}
\frac{\phi_1}{\phi_2}=\frac{\alpha_2\eta_1\dot{\gamma}+\frac{1}{T_0}H_2D_2^2}{\alpha_1\eta_2\dot{\gamma}+\frac{1}{T_0}H_1D_1^2}
\end{equation}

The  above estimate appears to be an extension  of   several  empirical formulas that can be found in the literature. For example, if we neglect the  elasticity of the two fluids  (i.e. we put $W_{1/2}=W_{2/1}=0$), or if the  mixing is very vigorous (i.e. if $\dot{\gamma}$ is large so that the terms in (\ref{result}) involving the viscosity are much larger than the terms involving the elastic energy),  or also if $T_0$ is large then (\ref{result}) becomes
$\frac{\phi_1}{\phi_2}=\frac{\alpha_2\eta_1}{\alpha_1\eta_2}$. If, in addition, we neglect the shape factor (i.e.  $\alpha_1=\alpha_2=1$)  then we arrive at the estimate $\frac{\phi_1}{\phi_2}=\frac{\eta_1}{\eta_2}$ which  is indeed the empirical formula introduced in \cite{phinv}.

\section{Concluding remarks}\label{CR}

Reducing dynamics $MESO\rightarrow meso$ is a dynamics bringing a mesoscopic level of description  (called \textit{MESO level}) to another mesoscopic level of description (called  \textit{meso level}) that involves less details. By identifying the reducing dynamics with thermodynamics we have been able to formulate a general  thermodynamics that encompasses the classical equilibrium thermodynamics (corresponding to $equilibrium\rightarrow equilibrium$), the equilibrium statistical mechanics (corresponding to $MICRO\rightarrow equilibrium$),  mesoscopic equilibrium thermodynamics (corresponding to $MESO\rightarrow equilibrium$), and thermodynamics of externally driven systems (corresponding to $MESO\rightarrow meso$). The general thermodynamics is presented in three postulates.

The first postulate (called Postulate 0 in order to keep as much as possible the traditional terminology) states that there exist well established mesoscopic levels of description. By well established we mean well tested with experimental observations. This postulate generalizes the  postulate of the existence of the equilibrium states that serves as a basis of the classical equilibrium and the Gibbs statistical equilibrium thermodynamics.

The second postulate (called Postulate I) is about state variables used on the \textit{MESO} and \textit{meso levels} and about potentials needed to formulate mechanics. Again, this postulate generalizes  the classical Postulate I in the classical equilibrium thermodynamics.

The third postulate (Postulate II) addresses  the process (called a preparation process) in which the macroscopic systems are prepared to states at which the \textit{meso} description is found to agree  with a certain family of experimental observations forming the experimental basis of the \textit{meso} description. The time evolution making the preparation process is called  reducing time evolution.
In the classical equilibrium thermodynamics this postulate  is the static Maximum Entropy principle (static MaxEnt principle) specifying only the final result of the preparation process.
In the general \textit{MESO} and  \textit{meso} descriptions, it is dynamic MaxEnt principle postulating  equation  governing the time evolution  making the preparation processes. Two important  results   arise on the \textit{meso level} from  the static  or the dynamic MaxEnt principles. First, it  is the \textit{meso level} time evolution (the time evolution reduced from the time evolution on the \textit{MESO level}). Second, it is
the  fundamental thermodynamic relation that  is constructed from  the potential generating the reducing time evolution. The generating potential has the physical interpretation of  entropy if the \textit{meso level} in the approach $MESO \rightarrow meso$ is the equilibrium level and entropy production (or related to it quantity) if the \textit{meso level} in the approach  $MESO \rightarrow meso$ is a general \textit{meso level}.
In the classical,   or the Gibbs statistical,  equilibrium thermodynamics (i.e. if the meso-level in the approach $MESO  \rightarrow meso$  is the equilibrium level) the reduced \textit{meso} dynamics is in this case no dynamics. The fundamental thermodynamic relation is, in the context of the classical  or the Gibbs statistical   equilibrium thermodynamics,  the classical equilibrium fundamental thermodynamic relation. In the context of \textit{MESO} and \textit{meso}  descriptions of externally driven macroscopic systems it is a new relation on the \textit{meso level} representing its thermodynamics. This thermodynamics is not directly related to the \textit{meso} dynamics (as, indeed,  the classical equilibrium fundamental thermodynamic relation is in no relation to no dynamics at equilibrium).
It represents an extra information about  macroscopic systems. The \textit{meso} dynamics is the \textit{MESO} dynamics seen on the \textit{meso level} and the thermodynamics is an information extracted from the way how the details (that are seen on the \textit{MESO level} but are invisible on the \textit{meso  level}) are being forgotten.

The fourth postulate (Postulate III of the classical equilibrium thermodynamics) addresses the value of entropy at zero absolute temperature. Investigations of macroscopic systems at such extreme conditions are outside the scope  of this paper. We are not  extending this postulate to mesoscopic dynamical theories.

In conclusion, we have demonstrated that if we limit ourselves to one fixed \textit{meso level} (e.g. the level of fluid mechanics) then the dynamic and the thermodynamic modeling  represent two essentially independent ways to investigate externally driven macroscopic systems. In particular, the validity and the pertinence of the thermodynamic \textit{meso} models does not depend on establishing their relation to the dynamic \textit{meso} models. If however we make  our investigation simultaneously on two well established levels, one \textit{MESO level} (that involves  more details than the \textit{meso level}) and the other the chosen \textit{meso level}  then we can derive both the \textit{meso}  dynamics and the \textit{meso} thermodynamics  from  the \textit{MESO}  dynamics. The derivation consists of splitting the \textit{MESO} time evolution into reducing time evolution (providing the \textit{meso} thermodynamics) and reduced time evolution that becomes the \textit{meso} time evolution. In most investigations of reductions the attention is payed only the the reduced dynamics. We hope that this paper will stimulate investigations in both reduced and reducing dynamics.

What are the arguments supporting the three postulates of the general thermodynamics? In the case of $equilibrium\rightarrow equilibrium$ passage they become the standard postulates of the classical equilibrium thermodynamics. In the case of $MICRO\rightarrow equilibrium$ passage they become a  formulation (equivalent to many  other existing formulations) of the Gibbs equilibrium statistical mechanics. In the case of $MESO\rightarrow equilibrium$ passage, a large body of supporting evidence has been collected (see in particular \cite{Grmadv} and references cited therein and in \cite{Obook}). In the case of $MESO\rightarrow meso$ passage there is much smaller number of examples that have been worked out. The support in this case comes, in addition to the support coming from the detailed analysis in the examples, from the unification  that the general thermodynamics brings (see more in  the text at the beginning of  Section \ref{EX}).
\\
\\

\textbf{Acknowledgements}

This research was partially supported by  the Natural Sciences and Engineering
Research Council of Canada. 
\\

\end{document}